\documentclass[traditabstract]{aa} 
\usepackage{txfonts}
\usepackage{graphicx}
\def\la{\raise.5ex\hbox{$<$}\kern-.8em\lower 1mm\hbox{$\sim$}}
\def\ma{\raise.5ex\hbox{$>$}\kern-.8em\lower 1mm\hbox{$\sim$}}

\def\kms{$\rm km\, s^{-1}$}
\def\cm3{$\rm cm^{-3}$}
\def\Ts{$\rm T_{*}$~}
\def\Vs{$\rm V_{s}$}
\def\n0{$\rm n_{0}$}
\def\B0{$\rm B_{0}$}

\def\Ne{$\rm N_{e}$~}
\def\Te{$\rm T_{e}$~}

\def\erg{$\rm erg\, cm^{-2}\, s^{-1}$}

\def\L12{L$_{12\mu m}$~}
\def\F12{F$_{12\mu m}$~}

\def\Hb{H$\beta$~}
\def\Hbb{H$\beta$}
\def\Ha{H$\alpha$}
\def\Hg{H$\gamma$}
\def\Hd{H$\delta$}

\def\Ly{Ly${\alpha}$~}

\def\RO3{R$_{[OIII]}$}

\begin{document}

   \title{Interpretations of galaxy spectra at high redshift
}

   \subtitle{The \Hg/\Hb excess}

   \author{M. Contini \inst{1,2}
}
   \institute{Dipartimento di Fisica e Astronomia, University of Padova, Vicolo dell'Osservatorio 3. I-35122 Padova, Italy
         \and
             School of Physics and Astronomy, Tel Aviv University, Tel Aviv 69978, Israel\\
}

   \date{Received }


  \abstract{
Spectra from the cosmic dawn obtained with JWST/NIRSpec (James Webb Space Telescope near-infrared spectroscopy) in the SMACS0723 Early Release 
Observations are now available. Analyses carried out by different teams indicate poor to extremely low oxygen metallicities (log(O/H)+12 $<$ 8.0), 
a characteristic feature of pristine galaxies.
In this work, we present new modelling of spectra emitted by objects in the redshift range 2.16 $\leq$ z $\leq$ 8.68,  
 including their recently corrected spectra in the z = 2-9 range. 
The models account for both photoionisation and shock processes.
Our aim is to identify similarities and differences with respect to local galaxies by searching for possible remnants of pristine galaxies among 
low-z objects.  We analyse selected emission-line ratios in relation to elemental abundances and physical parameters.
We find that the gaseous clouds within galaxies at cosmic dawn have preshock densities at least a factor of 100 higher than those in local galaxies, 
but comparable to those calculated for local metal-poor galaxies at 0.005 $<$ z $<$ 0.05. The metallicities log(O/H)+12 and log(Ne/H)+12 range 
between 7.9 and 8.55, and between 7.0 and 7.48, respectively.
Uncorrected observed  \Hg/\Hb line ratios  are mostly $>$ 0.5, indicating high  temperatures ($>$ 10$^5$ K) in the emitting gas. 
Clear affinities are evident between the
high-z galaxy spectra and those of local metal-poor galaxies at 0.005 $<$ z $<$ 0.05.
However, in order to reproduce all the observed line ratios for each spectrum - including \Hg/\Hb values as high as $\sim$ 0.8 - emission from 
cloud fragments was added
to that from the main clouds in the pluri-cloud models. We suggest that fragments close to pristine galaxies were destroyed by events that occurred 
between z $>$ 6 and z $<$ 0.05, whereas some cloud remnants of pristine galaxies survived and are now found embedded, for example, 
at 0.005 $<$ z $<$ 0.05.
}




\maketitle

\section{Introduction}

JWST/NIRSpec spectroscopy from the Early Release Observations of the galaxy cluster SMACS J0723.3-327 has now been released, providing detections of nebular emission 
lines, among others, in spectra at z $>$ 5 (Pontoppidan et al. 2022; Curti et al. 2023; Trump et al. 2023; Topping et al. 2024; Schaerer et al. 2022; 
etc.). 
In particular, Pontoppidan et al. (2022) were responsible for the design and execution of the observations.

In recent years, spectral analyses have focused mainly on the physical and chemical parameters of galaxies at low redshift (Nakajima et al. 2022; Contini 2026b);
 Berg et al. 2016; Marino et al. 2013; Niino et al. 2016; etc.). JWST observations allow a direct comparison of physical conditions over 13 Gyr of 
 cosmic
time, using the same set of rest-frame optical emission lines from cosmic dawn to the present epoch (Trump et al. 2023). The redshift range accessible 
to galaxy spectral investigations, previously limited to z $\leq$ 4, has now been extended to z $>$ 5.

Spectral lines are essential for constraining the modelling process. The number and variety of lines provided by JWST observations are now 
sufficient to describe 
quantitatively most of the physical conditions in galaxies. From data analyses, the authors agree that most low-z starburst (SB) galaxies are 
oxygen-poor. This 
characteristic is typical of pristine  (so-called ‘early’) galaxies, i.e. near-primordial stellar systems composed almost entirely of hydrogen 
and helium, with 
very few heavy elements.

We  wonder whether low oxygen abundance alone is sufficient to identify galaxies that have survived from cosmic dawn to low redshift; this is not the sole motivation 
for the present study.  Our aim is to broaden the investigation of galaxy physical conditions across different redshifts.
We proceed by applying the detailed modelling method to each spectrum in the surveys.  After determining the characteristic sets of physical and chemical parameters, 
we investigate whether remnants of cosmic-dawn galaxies can be identified at low redshift, at the end of their evolutionary path, in the light of their nucleosynthesis 
origin, ${\alpha}$-capture processes, the CNO cycle, and interactions with strong Wolf-Rayet stellar winds, etc. During the step-by-step modelling process, we identified 
several interesting issues, for example that models accounting for both photoionisation and shocks may lead to pluri-cloud solutions.

Recently, spectra of Extreme Metal-Poor Galaxies (EMPGs) presented by Nakajima et al. (2022) in the redshift range 0.00574-0.05368 were modelled by Contini (2026b) 
taking into account the coupled effects of photoionisation and shocks. Very low oxygen metallicities had previously been derived using the strong-line method 
(Izotov 2006). The choice of shock-inclusive models was discussed in detail by Calabro' et al. (2023), who examined whether line ratios calculated by shock-dominated (SD) 
models could reproduce high-z galaxy spectra. They excluded shocks as the dominant contributors to the global emission, based on their interpretation of active galactic 
nuclei (AGN) and SB galaxy samples at low redshift.

We note, however, that shocks can heat the gas to high temperatures depending on the shock velocity. They allow the emission of high-ionisation lines and, in some 
cases, it has been found that highly ionised iron coronal lines and intermediate-ionisation oxygen lines can coexist in the same spectrum (Fonseca-Faria et al. 2023). 
We will also investigate the unusually high \Hg/\Hb values ($\geq$ 0.8) reported following an unexpected observational incident affecting the Trump et al. ID4590 galaxy 
during visit 7. This issue will be discussed using the most appropriate modelling approach.

The metallicities obtained by Contini (2026a) using the code {\sc suma} (Contini \& Viegas 2001a,b and references therein) for low-z galaxy spectra show oxygen depletions 
smaller by factors of 2-3 than those derived using the strong-line method, and O/H lower than solar by factors $\leq$ 5 for 0.00574 $<$ z $<$ 0.0537 EMPGs (Contini 2026b),
 while N/H is lower than solar by factors $\leq$ 10. We adopt the following solar abundances (Grevesse 2019):
(He/H)${\odot}$ = 0.085,
(C/H)${\odot}$ = 2.75$\times$10$^{-4}$,
(N/H)${\odot}$ = 6.76$\times$10$^{-5}$,
(O/H)${\odot}$ = 4.9$\times$10$^{-4}$,
(Ne/H)${\odot}$ = 1.2$\times$10$^{-4}$,
(Mg/H)${\odot}$ = 3.63$\times$10$^{-5}$,
(Si/H)${\odot}$ = 3.2$\times$10$^{-5}$,
(S/H)${\odot}$ = 2$\times$10$^{-5}$,
(Fe/H)${\odot}$ = 3.0$\times$10$^{-5}$.
The value (Ne/H)${\odot}$ = 1.2$\times$10$^{-4}$ was updated by Young (2018). The preshock densities were found to be higher by a factor $>$ 10 
in EMPGs than in other local galaxies.

In spectra of high-z objects observed by Trump et al. (2023), optical lines such as [OIII]5007+4959, [OIII]4363, [OII]3727+3729 and [NeIII]3869 (hereafter [OIII]5007+, 
[OIII]4363, [OII]3727+, and [NeIII], respectively), together with the Balmer lines \Hg, \Hd\ and \Hb, are reported in their Table A.1.  Curti et al. 
(2022, 2023) 
suggested that one of the shutters for ID4590 (visit 7) did not open. This is supported by the strong discrepancy in the \Hg/\Hb ratio between 
visits 7 and 8.

High \Hg/\Hb ratios ($\geq$ 0.8) originate from hot gas. Although the modelling results for the Trump et al. visit 7 spectrum are not entirely reliable, we have given 
particular attention to the unusual \Hg/\Hb = 0.84 ratio, as similar values have been reported in the literature (e.g. de Ugarte Postigo et al. 2014). Other spectra 
with \Hg/\Hb $\geq$ 0.7 were published by Sanders et al. (2024), based on  flux-calibrated line ratios, and by other authors.

We considered whether the high \Hg/\Hb ratio could be explained without invoking strong correction factors. High \Hg/\Hb values (from $>$ 0.5 up 
to $\geq$ 0.8) in 
spectra dominated by strong optical lines provide an opportunity to apply pluri-cloud models to reproduce the entire spectrum. In this approach, the weighted sum of 
line fluxes calculated under ‘normal’ Osterbrock physical conditions is combined with line fluxes derived from shock-dominated models. The shock velocity determines the 
maximum gas temperature and the spatial extent of the high-temperature region, which must be chosen to remain smaller than the distance to the recombination temperature 
drop (see Fig. 1). The debris represent residual fragments resulting from cloud fragmentation caused by turbulence at the shock front.  
We refer to this matter-bounded debris as ‘fragments’ (fr).

In such fragment spectra, strong optical lines are close to zero, while \Hg/\Hb ratios remain relatively high. These fragments preserve the physical and chemical properties of 
the original clouds, which were disrupted by explosions or by fragmentation depending on the surrounding conditions through  which they propagate.

The significant lines reported by Curti et al. (2023) for ID4590, ID6355 and ID10612 in the optical-near-UV range, and the HeI5876, HeII4686, [NII]6584 and [SII]6716,
6730 lines reported by Topping et al. (2024) for RXCJ2248 at z = 6.1, are reproduced by our models. Previous estimates of the physical and chemical conditions for these 
spectra were obtained using the strong-line method. We instead apply detailed modelling, taking advantage of the relatively large number of 
emission lines from different ions.

The  flux-calibrated spectra presented by Sanders et al. (2024), based on JWST/NIRSpec observations in the redshift range 2-9, are also modelled. 
These data help to  shape
the parameter trends towards high z and to explore N/O ratios as a function of metallicity. To model high-redshift galaxy line ratios, we adopt 
the code {\sc suma}, previously applied to different types of objects across a range of redshifts.

The modelling method is briefly described in Sect. 2. The results are compared with the data in Sect. 3. 
We begin presenting the \Hg/\Hb line ratio question.
The high \Hg/\Hb $\sim$ 0.8 and \Hg/\Hb $>$ 0.5 values are explained in Sect. 3.1.
Then, the optical emission-line flux ratios for 
galaxies at z $>$ 5 reported by Trump et al., based on JWST/NIRSpec spectroscopy, including the uncorrected data in order to test pluri-cloud models, 
are analysed.
 We revisit 
their spectra, particularly to discuss the \Hg/\Hb ratio. 

In Sect. 3.2 we use the ID4590 spectrum observed during visit 8 as a representative case, as it includes both relatively high \Hg/\Hb ratios and strong optical lines, 
to illustrate pluri-cloud modelling. Sect. 
3.3 completes the modelling of the Trump et al. spectra. Trump et al. suggested that the correction method adopted by Curti et 
al. (2022, 2023) was the most appropriate; the  flux-calibrated spectra are discussed in Sect. 3.4.

We report the UV-optical data presented by Topping et al. (2024) for RXCJ2248 at z = 6.1 and compare them with our model results in Sect. 3.5. 
The Sanders et al. 
(2024) galaxy flux-calibrated sample further enriches our analysis of gas physical conditions and relative abundances in the redshift range 2-9, 
particularly at z $>$ 3. The modelling of their spectra is presented in Sect. 3.6.

Observers have suggested that ‘the high ionisation state of z $\geq$ 6 HII regions likely reflects conditions associated with powerful bursts 
that appear common in early 
galaxies.’ We also consider other galaxy types. In Sect. 4 we discuss the trends of \Vs\ and \n0\ for galaxies in the redshift range 0.0001-9, 
and the behaviour of N/O as a function of redshift and metallicity, based on models that account for the combined effects of shocks and 
photoionisation.

\section{Models}

The emitting clouds are divided downstream into multiple slabs (up to a maximum of 300), with geometrical thickness  calculated automatically 
in order to obtain a smooth temperature trend  throughout the emitting nebula.
Electron temperatures and electron densities (\Te\ and \Ne, respectively) are calculated in each slab, together with bremsstrahlung and 
line-flux intensity increments through radiative transfer. These contributions are then summed over the full set of slabs.

Initial estimates of \Te\ and \Ne\ are obtained from observed line ratios, such as [OII]3727/[OII]3729 and [SII]6716/[SII]6730 for density diagnostics, and 
[OIII]5007+/[OIII]4363 for temperature diagnostics. However, the [OII] doublet is often blended, and the [SII] ratios can be affected by ISM abundances.
Immediately downstream of the shock front, the gas may reach very high temperatures due to collisional processes. \Te\ and \Ne\ are computed 
following the method of Cox (1972).
The preshock density increases across the shock front through the adiabatic jump (by a factor of 4). The gas is subsequently compressed downstream, 
depending on the temperature, the ram pressure, and the magnetic field. Compression in each slab is calculated using the  combination of 
Rankine-Hugoniot equations (Cox 1972) for the conservation of mass and momentum.
In our models, compression operates both in a shock-dominated regime and in a composite shock + photoionisation regime, for example in AGN and SB 
galaxies.  The gas temperature in each downstream slab is determined by balancing collisional heating due to the shock, heating from the photoionising 
radiation flux, and cooling via free-free, free-bound, and line emission. Consequently, \Ne\ and \Te\ downstream do not follow a uniform trend 
(Figs. 1 and 2).

The thermal balance procedure is automatically iterated until equilibrium is reached in each slab, after which the calculations proceed to the 
next slab. 
The calculation stops when all ions have recombined in radiation-bounded models, or when the cloud is truncated in matter-bounded cases. 
All dominant line ratios 
and the continuum are calculated throughout the cloud for each model, including lines not observed, since they contribute to the gas cooling rate.
The calculated spectrum is compared with the observations, taking into account that strong-line fluxes typically have uncertainties of 10-20\%
while weak lines may
 have uncertainties up to 50\%. The computational uncertainty is approximately 10-20\%.

A two-phase gas regime often develops within the clouds (see Fig. 1), because the cooling rate via line emission in the post-shock region increases 
sharply as the 
temperature decreases from the shock front towards $\sim$10$^5$ K. In this regime, far-UV to optical lines and near-infrared emission accelerate 
the recombination process.

In summary, the input parameters defining a model are as follows:
\Vs\ is the shock velocity;
\n0\ is the preshock gas density;
$D$ is the geometrical thickness of the emitting cloud;
$U$ is the ionisation parameter;
\Ts\ is the SB effective temperature;
$F$ is the photoionising power-law flux from AGN at the Lyman limit.
The parameter str distinguishes between accretion towards the radiation source (str = 0) and outward 
ejection (str = 1).
 We cannot decide  whether the clouds corresponding to  pure SD models are  accreted towards the radiation source or  ejected outwards from the galaxy
because there is a single external ionizing source (the shock).
A double source (shock and photoionization) indicates accretion  or ejection  if the fluxes from the two sources reach the same
cloud edge  (Fig. 1) or  opposite edges (Fig. 2), respectively.

Unless otherwise specified, all models adopt a preshock magnetic field of \B0 = 10$^{-4}$ Gauss. A strong magnetic field can reduce the degree 
of compression. We consider the full set of emission lines in each spectrum.

\section{Modelling the spectra}

In the present investigation the O/H and Ne/H  relative abundances  are  calculated from  the Trump et al. observed  spectra
 \footnote{The JWST data used by Trump et al. (2023) can be found in MAST at DOI:10.17909/67ft-nb86.}
The He/H, C/H, N/H, O/H and Ne/H ratios resulted  by modelling the Curti et al. spectra, while 
C/H, N/H, O/H  and Mg/H abundance ratios derived from Topping et al. line ratios. From Sanders et al. data  we calculate  N/H, O/H  and Ne/H 
ratios.  Sulphur lines were missing or approximated for most  objects.
In all surveys \Hb  and \Hg~ observed flux were reported.
The lines  corresponding to the other elements (Si, Cl, Ar and Fe), included in the code but not provided by the observations,  
were calculated  with solar abundances. 
 The different methods  used by  
Curti et al. (2023), Schaerer et al. (2022), Rhoads et al. (2022) and Trump et al.  are summarized by Taylor et al. (2023).

\subsection{\Hg/\Hb}

The unfortunate event that occurred to  ID4590 observation  during visit 7 lead us to investigate the high \Hg/\Hb line ratios. 
The results brought us to develop the pluri-cloud method in the modelling process of 0.5 $\leq$ \Hg/\Hb $<$0.8.

The observed lines in the present sample of galaxy spectra cover the optical + near-UV range. At those wavelengths, the calculated \Hg/\Hb ratios
  are  $\sim$ 0.42-0.49  adopting the normal physical  conditions of the emitting nebulae indicated
by Osterbrock (1974)  e.g. by '\Te$\sim$10000 K and a large range of densities'. 
It is difficult  for pure photoionization models which heat the gas to a  maximum of 2-3$\times$10$^4$K to achieve 'abnormal'  high
\Hg/\Hb line ratios within spectra  which contain 'normal' line  in the optical - near-UV range.
Therefore, \Hg/\Hb  as high as  $\sim$0.8  can be  reached by high-velocity  shocks  which  yield high temperatures.

  Some extreme  cases (\Hg/\Hb$\leq$0.8) were found  e.g.
for the de Ugarte Postigo et al. (2014, their Table A.5, FORS OT site) spectrum of
the short gamma-ray burst (SGRB)  30603B host galaxy at z$\sim$ 0.3565$\pm$0.0002. \Hg/\Hb resulted $\sim$0.77.
Sanders et al. (2020 in their Table A.1)  for
 the  flux-calibrated sample  observed from z=1.5-3.5  galaxies by COSMOS 23895  reported \Hg/\Hb=0.8.
  Other examples are given by Sanders et al (2024) for
 the  calibrated fluxes observed from galaxies (e.g. ID792) in the  z range 2 - 9 (Table A.8) with \Hg/\Hb=0.7.

The  model which   eventually  reproduces \Hg/\Hb$\sim$0.7  is m(fr)* (Table A.1) where m is for model,
   fr for fragment and the asterix  indicates  calculation  results  for  high \Hg/\Hb.
m(fr)*   is  calculated with \Vs=8300 \kms   which corresponds to T= 10$^9$K
(T=1.5$\times$10$^5$ (\Vs/100\kms)$^2$),  for a  SD model.   
This  temperature is  recommended  as the upper limit of  super luminous supernova (SLSN).
SLSN are  present at cosmic dawn  (Tanaka, Takashi \& Yoshida 2013).

 A high preshock density \n0=10$^7$ \cm3 is adopted to accelerate the  plasma recombination,  reducing the geometrical 
 thickness of the nebula to  about those of a fragment.
  Model m(fr)* appears in Table A.1  and A.2 as a general example.
 The  lines  emerging  in the UV, optical and near-IR ranges calculated by this model at  such high temperatures
  correspond to HeII4686/\Hb=0.40, HeII1640/\Hb=8.2, \Ly/\Hb= 92.19 and
 Fe coronal line ratios [FeX]/\Hb=3.05, [FeXI]/\Hb=0.62 and
[FeXIII]/\Hb=0.3  from Topping et al. and  all the
 line ratios  which are strong enough to be observed adopting cosmic abundances.

Among Trump et al., Topping et al.  and Sanders et al.   observations  most of them show
\Hg/\Hb  $\geq$0.5.  Those ratios are emitted from  warm gas (T$>$5$\times$10$^4$K).
The other line ratios included in  the spectra, on the other hand, are  easily reproduced by models adopting
the usual values of \Te, \Ne  and of the parameters referring to the photoionizing flux.
Such \Hg/\Hb ratios are  incompatible with the  conditions of  gas emitting the strong optical lines.
The  analysis of  several spectra  reported in the following, therefore, suggests  to use  pluri-cloud  models which result
from the sum of at least two  different ones.
One model  is selected among  the  shock dominated fragments  with high \Vs and the  other  is
  radiation dominated   reproducing the observed  optical-UV line ratios.
In this case, to  obtain a high  temperature region  in  the fragment, the geometrical thickness, adopted for the cloud, 
is reduced by interrumping $D$  at the right distance from the shock-front  (Fig. 1). We will use pluri-cloud models for all the spectra  with \Hg/\Hb $>$ 0.5.

 Another approach by Cox \& Mathews (1969, Fig.1)  should be mentioned
for large optical thickness of the \Ly line because the increase of the \Ha~ optical thickness
 leads to an increase of both
 \Ha/\Hb and \Hg/\Hb. From Contini (2003, Fig.1) we notice that \Hg/\Hb can
 reach 0.63 by appropriated  values for \Vs~ and \B0.
 They are  not sufficient to  explain the \Hg/\Hb$\sim$0.8 ratios. Moreover,
the self-absorption models  in general are limited to  homogeneous statical isothermal regimes,  therefore, they will be neglected
in the following interpretation.

\begin{figure}
\includegraphics[width= 8.70cm]{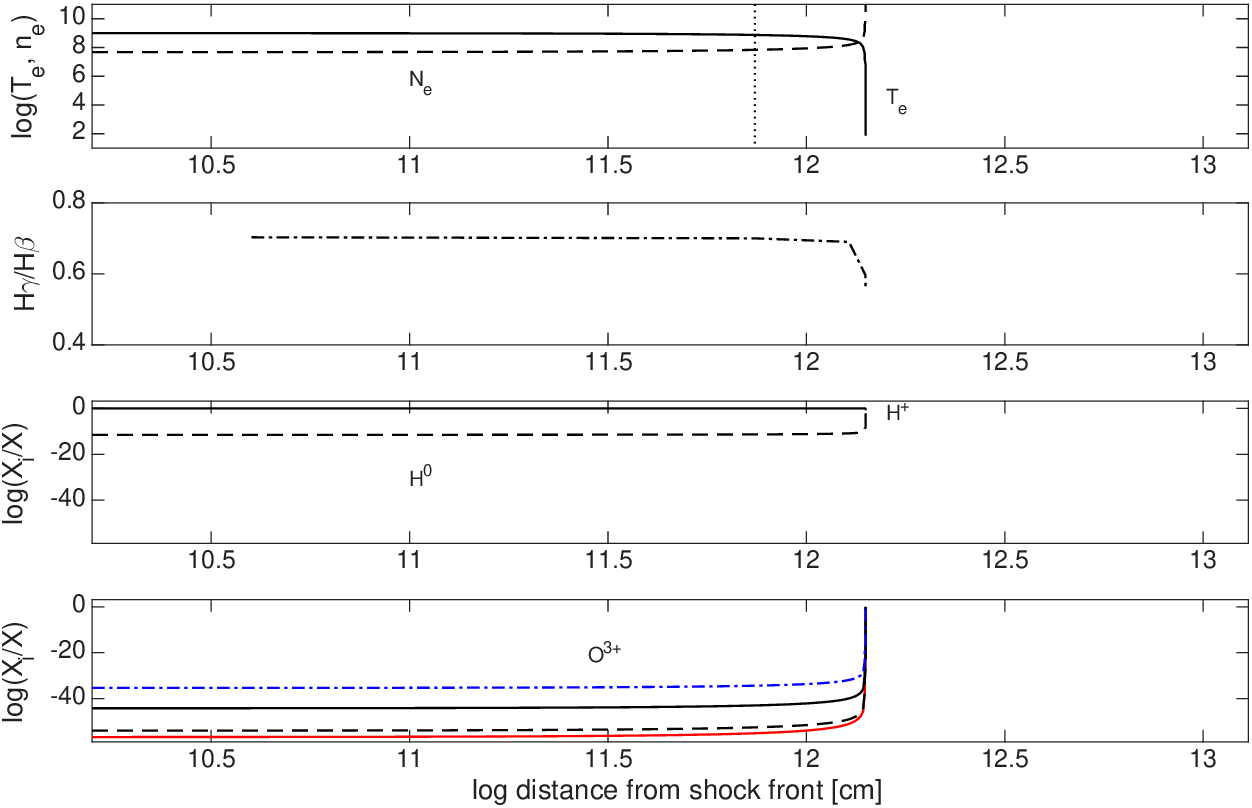}
\includegraphics[width= 8.70cm]{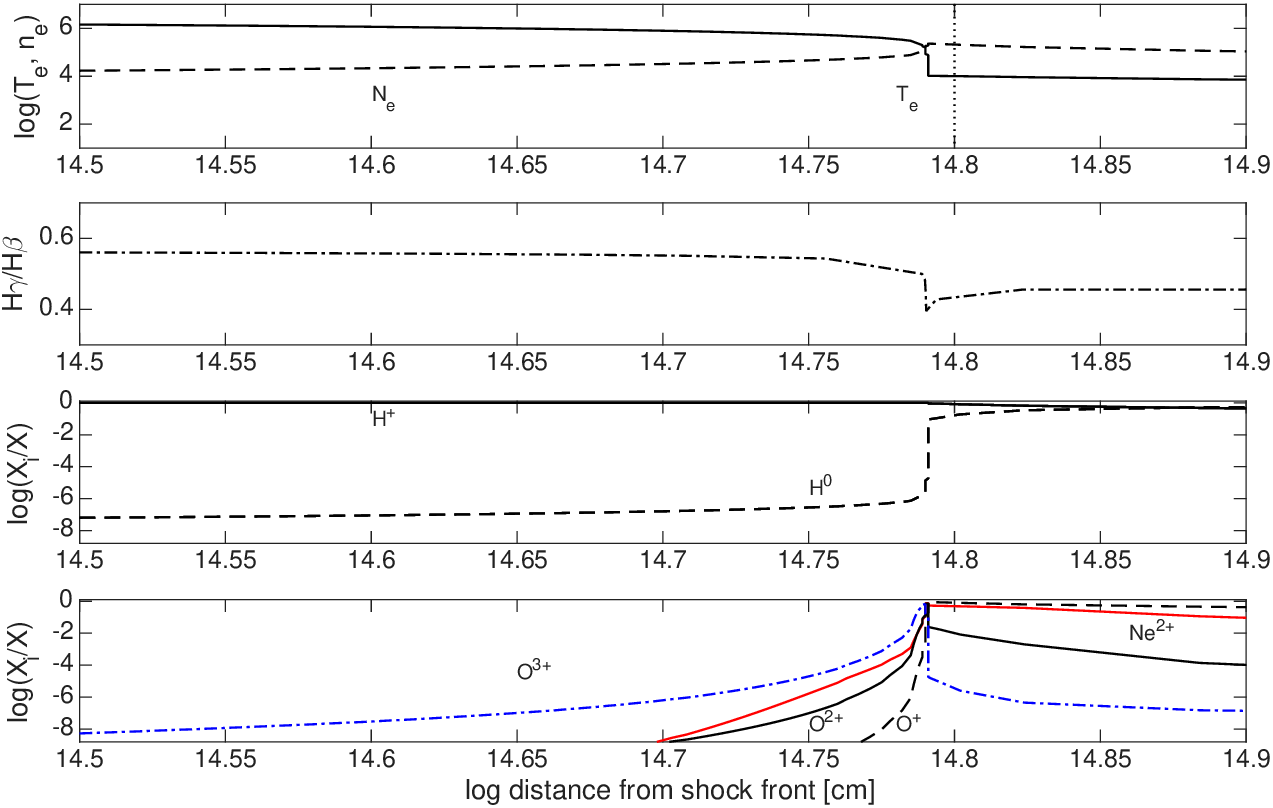}
\caption{Top diagrams: profiles of the electron temperature \Te and of the electron density  \Ne,
of \Hg/\Hb line ratio and of the fractional abundance of H ions
and of O$^+$/O, O$^{2+}$/O (black), O$^{3+}$/O (blue) and  of Ne$^{2+}$/Ne (red)  ions  throughout
SD fragment model m(fr)*.
Bottom: the same for the accretion (str=0) model mtr$_{8.49}$(cl8A) (shock+AGN).
Black vertical dotted lines show the edges of the matter-bounded downstream regions.
}
\end{figure}

\subsection{Trump et al. on ID4590 visit 8}

We have attempted to model the Trump et al. observed spectra in detail, taking into account that the applied corrections appear to be relatively small. The data from
Trump et al. are reported in their Table A.1. The authors state that ‘the spectra have reliable relative spectrometry for pairs of emission lines that are closely
separated in wavelength’, but they avoid analyses and interpretations that require absolute spectrophotometric calibration.
 However, the relatively uniform \Hg/\Hb ratios, ranging between 0.43 and 0.48, suggest that dust effects are modest and that strong corrections
are unlikely to be necessary. On the other hand, the presence of unusually high ratios indicates that pluri-cloud models may provide an efficient 
explanation.

The results of modelling the Trump et al. line ratios are   presented in Tables A.1 and A.2.
The panels in Fig. 1  show  the  \Te, \Ne, \Hg/\Hb profiles, the fractional abundance
of H$^0$/H and H$^+$/H, of  O$^+$/O, O$^{2+}$/O, O$^{3+}$/O and  Ne$^{2+}$/Ne    through the fragment corresponding
to the line domains of \Hb, [OII], [OIII], [OIV] and [NeIII] for model m(fr)* (top)  and for model mtr$_{8.49}$(cl8A) through the cloud (bottom), respectively.

The  spectra reported by  Trump et al.  in their Table A.1  show
\Hg/\Hbb~ ranging between 0.54 and 0.59 in 6  of the 10 spectra.
 It is  important to  maintain the nice  fit  of all the other  lines calculated  for the cloud.
In this case, to  obtain a high  temperature region  in a fragment, the geometrical thickness $D$ is  interrupted at a distance
 from the shock-front  before  gas recombination (Fig. 1).
Galaxies in the Nakajima et al. low-z sample show 0.443 $\leq$ \Hg/\Hb\ $\leq$ 0.5, with a single maximum of 0.53 for HSCJ1411–0032. This suggests that, apart from this
outlier, the shock parameters derived for the Nakajima sample galaxies (Contini 2026b, Fig. 8, left panels) are mutually similar and systematically different from those 
obtained for the high-z Trump et al. objects. Therefore, the high-redshift galaxy spectra differ from those of low-z EMPGs, most likely owing to a common physical 
event. We suggest that fragments present near the emitting clouds in high-z ($>$ 6) galaxies did not survive to low redshift (z $<$ 0.05).

For ID4590 (visit 8), the Trump et al. spectrum can be reproduced either by an SB model, mtr${8.49}$(cl8), (m is for model, tr for Trump et al,
8.49 is for z and cl8 for cloud on visit 8) or by an AGN model, mtr${8.49}$(cl8A), with input parameters \Vs, \n0, and $D$  within the observational 
uncertainties. An AGN matter-bounded model with \Vs = 400 \kms, \n0 = 2000 \cm3, and a nuclear flux lower than typically adopted for an AGN 
narrow-line region (NLR) reproduces the observed line ratios at a distance $D$ = 1.4$\times$10$^{-4}$ pc from the shock front, except for \Hg/\Hb.
An SB model with \Ts = 5$\times$10$^4$ K, $U$ = 0.006, and $D$ = 1.8$\times$10$^{-4}$ pc also provides a satisfactory fit, depending on the 
weight assigned to the
[OII]/\Hb\ ratio. A fragment model with similar parameters but $D$ = 10$^{-5}$ pc (Table A.1, model mtr${8.49}$(fr8)) is added to the cloud component 
with a relative 
weight W${fr}$/W${cl}$ $\leq$ 11. The relatively high W${fr}$/W$_{cl}$ ratio compensates for the low \Hb\ flux emitted by the fragment.
We adopt the SB model (mtr${8.49}$(cl8)), which reproduces all line ratios well (slightly less accurately for [OII]/\Hb), since a starburst scenario 
is more consistent 
with cosmic-dawn conditions than an AGN-formation process. For visit 8, a higher shock velocity is not required; instead, a matter-bounded model 
similar to mtr${8.49}$(cl8),
truncated at a suitably  reduced distance $D$ downstream of the shock front, yields a slightly warmer gas. The geometrically thin fragments 
($D \sim 2.5\times10^{-7}$ pc) share the same chemical composition as the main clouds.  For ID4590 visit 8, the O/H abundance is low, 
with 12 + log(O/H) = 7.84, lower than values found in metal-poor galaxies of the Local Group (Contini 2026b).

\subsection{ID6355, ID10612, ID5144 and ID8140 spectra (Trump et al., visits 7 and 8)}

The differences reported by Trump et al. between visits 7 and 8 for spectra at z = 7.6651, 7.6597, 6.3792 and 5.2753 fall within the observational 
uncertainties. 
The calculated \Hg/\Hb\ ratios that reproduce the Trump et al. observations are shown in Table A.1. These ratios are broadly similar across the spectra, 
except for ID8140 at z = 5.2753. The clouds at different redshifts require different combinations of \Vs, \n0, radiation sources, geometrical thickness
$D$, and elemental abundances in order to reproduce all observed line ratios. Between z = 8.4957 and z = 6.3792, the shock parameters do not change 
dramatically. At lower z (e.g. ID8140 at z = 5.2753), the preshock densities approach typical values for AGN narrow-line regions.

For {\it ID6355} at z = 7.6651, \Hg/\Hb\ = 0.59 (visit 7) and 0.48 (visit 8). For visit 7, a fragment component must be added to model mtr${7.66}$(cl7),
reproduced by an AGN model. The fragment model mtr${7.66}$(fr7) is combined with the cloud component using weights W${fr}$/W${cl}$ $\sim$ 13. 
The fragments appear to share the same chemical composition as the clouds. O/H is 0.73 solar and Ne/H $\sim$ 0.1 solar. On visit 8, the fragment 
component is no longer required, and O/H decreases slightly (by $\sim$ 5\%) within the observational uncertainty. 

For {\it ID10612} at z = 7.6597, visits 7 and 8 are reproduced by similar SB models (mtr${7.65}$(cl7) and mtr${7.65}$(fr7)), with O/H approximately half solar 
and Ne/H = 0.22 solar. The model includes a relatively low preshock magnetic field, leading to higher downstream densities despite an already high \n0.

For {\it ID5144} at z = 6.3792, the visit 7 spectrum (model mtr$_{6.37}$(cl7)) is consistent with an AGN, with a typical nuclear radiation flux and shock parameters 
similar to those of other Trump et al. objects. This spectrum resembles that of SDSS J1044+0353 (Nakajima sample, analysed by Contini 2026b) at z = 0.01317, modelled 
in the accretion scenario. On visit 8, an SB model reproduces the [OIII]4363/\Hb\ ratio with high precision. A fragment model with \Vs = 400 \kms\ is required on 
visit 8, slightly higher than the cloud velocity. 12 + log(O/H) decreases from 8.56 (visit 7) to $<$ 8.4 (visit 8).  The presence of two different photoionisation 
sources in the same galaxy at closely spaced epochs is unlikely and highlights the difficulty of model selection when the number of diagnostic lines is limited.

For {\it ID8140} at z = 5.2753, the calculated \Hg/\Hb\ ratios are unusually low: 0.13 (visit 7) and 0.23 (visit 8). A pluri-cloud model, combining two cloud components,
is required. If the observed \Hg\ and 2/3 [OIII]4363 fluxes on visit 8 are blended, \Hg/\Hb\ reaches 0.43, consistent with the SB model mtr$_{5.27}$(cl8). On visit 7, 
\Hg/\Hb\ is approximately half the visit 8 value and [OIII]4363 is completely obscured. The maximum O/H is 0.46 solar (visit 8).

\subsection{Curti et al. (2023)  flux-calibrated spectra: ID4590, ID6355 and ID10612} 

To address the anomalously high \Hg/[OIII]4363 ratio in the Trump et al. data, we analysed the spectra of ID4590, ID6355 and ID10612 flux-calibrated by 
Curti et al. (2023). The results are presented in Tables A.3 and A.4. Curti et al. suspected that one shutter did not open for ID4590 on visit 7 and 
therefore based their analysis on visit 8. Their correction procedure restores the classical Osterbrock \Hg/\Hb\ values typical of  Nakajima et al. 
galaxy spectra. 
The  flux-calibrated ratios change from 0.59 to 0.42 (ID4590), from 0.54 to 0.46 (ID6355), and from 0.55 to 0.45 (ID10612), consistent with 
local EMPGs.

A local analogue of ID4590 (corrected visit 8) is J082555 dwarf galaxy HII region (Berg et al. 2016), after reddening correction.  
Its line ratios were modelled by Contini (2026a, model B1m). 
The corrected UV line ratios of ID4590 are reproduced within a factor of two by a similar model. The Curti flux-calibrated spectrum of ID4590 is best 
fitted by model 
mcr$_{8.49}$(cl) (Tables A.3 and A.4), based on Contini (2026b). Small but significant discrepancies remain (e.g. for [OIII]4363/\Hb\ and [OIII]5007/\Hb). 
Ne/H is reduced by a factor of 2-3, O/H by 1-3 relative to solar, and C/H by a factor of 2 to reproduce CIV/\Hb\ and CIII]/\Hb, while N/H is less 
depleted.
However, the preshock density and shock velocity in model B1m are lower by factors of 27.5 and 8, respectively, compared with those in mcr$_{8.49}$(cl).
ID6355 and ID10612 are reproduced by models mcr$_{7.66}$(cl) and mcr$_{7.65}$(cl), respectively. These models are less representative of dwarf 
galaxy HII regions but are consistent with EMPGs.

The resemblance of the spectra between the z=0.003 galaxy J082555 and the z=8.64 galaxy ID4590 may suggest a pristine origin for J082555, given that 
the Curti  flux-calibrated spectrum results from Trump et al. high-z observations. For ID6355, an AGN-dominated model provides the best fit.

\subsection{Topping et al. (2024) on the RXCJ2248-ID spectrum at z=6.1}

Tables A.5-A.7 and Fig. 2 present the modelling of the spectrum reported by Topping et al. (2024). The main characteristic of this spectrum is the very 
high [OIII]5007+ / [OII]3727+ line ratio. This can be explained by a composite SD + SB radiation model with intermediate \Vs\ and a relatively 
high preshock density \n0, in the ejection case.
All emission lines are reproduced by adopting reduced elemental abundances, particularly for helium. He/H is reduced by a factor of 33 relative to 
solar, C/H by a factor of 4, N/H by 1.11, O/H by 2.5, Ne/H by 6.25, and Mg/H by 162. Elemental abundances are constrained when the element is observed 
in at least two different ionic species. We agree with Topping et al. that nitrogen is the least depleted element.

The very low Mg/H value (Table A.7) suggests that magnesium is locked into dust grains together with silicon, possibly in olivine-type silicates, 
(Mg,Fe)$_2$SiO$_4$.
Silicate grains are destroyed at shock velocities $\geq$ 200 \kms, which are relevant for the Topping et al. models of RXCJ2248 (Table A.7). 
At lower \Vs, Mg, Fe, Si  and O
remain incorporated in grains and are therefore removed from the gaseous phase. The presence of dust could be further confirmed if silicon lines are 
found to be weak, 
for example [SiVII]2168 in the UV or [SiVII]64921 in the infrared.
The C/H and He/H abundances are both significantly below solar values, whereas N/H is approximately solar. This pattern may indicate 
enrichment by Wolf-Rayet 
stellar winds, possibly analogous to that inferred for J082555.

The fragments, which share the same chemical composition as the main clouds, are geometrically thin. They compensate the cloud value 
of \Hg/\Hb = 0.41 by contributing a fragment component with \Hg/\Hb = 0.509, corresponding to W${fr}$/W${cl}$ $\sim$ 5.

The profiles of the main physical parameters throughout the Topping et al. cloud are shown in Fig. 2 for the configuration in which the shock 
and radiation act on opposite edges (ejection case, str = 1).

\begin{figure*}
	\sidecaption
\includegraphics[width= 6.0cm]{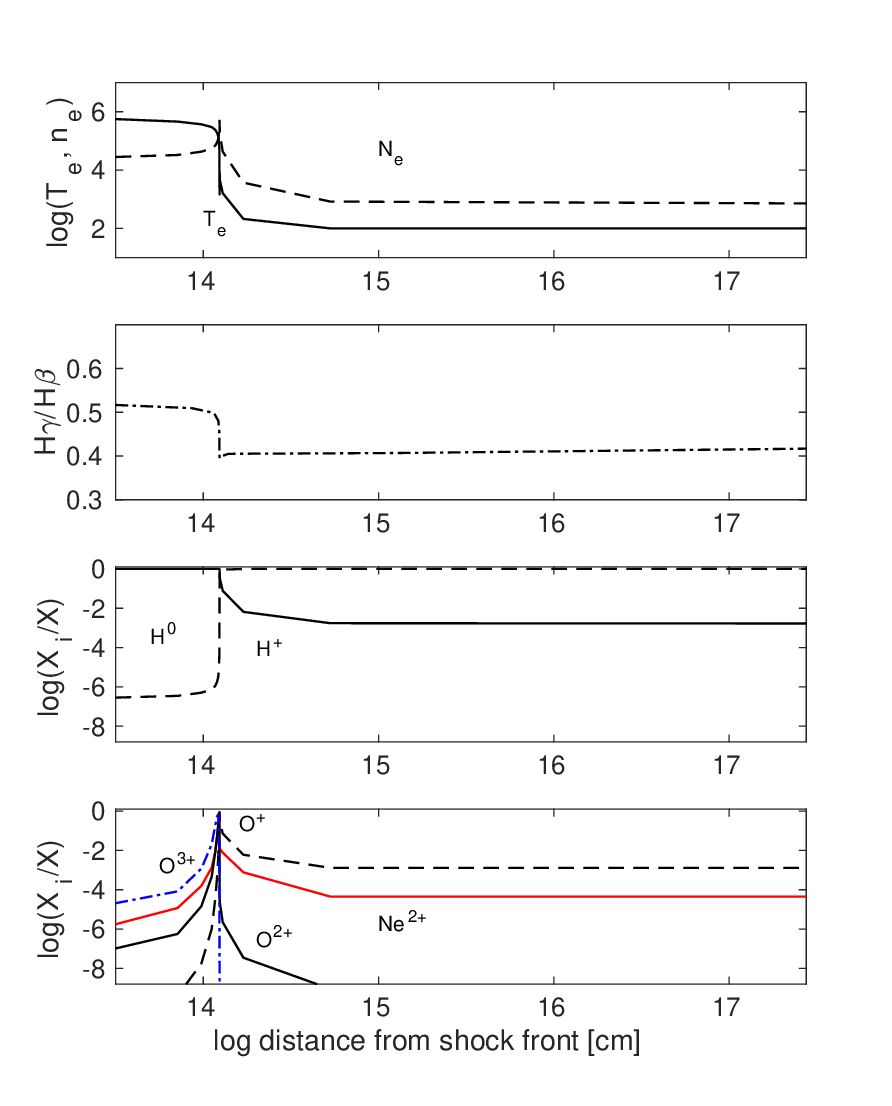}
\includegraphics[width= 6.0cm]{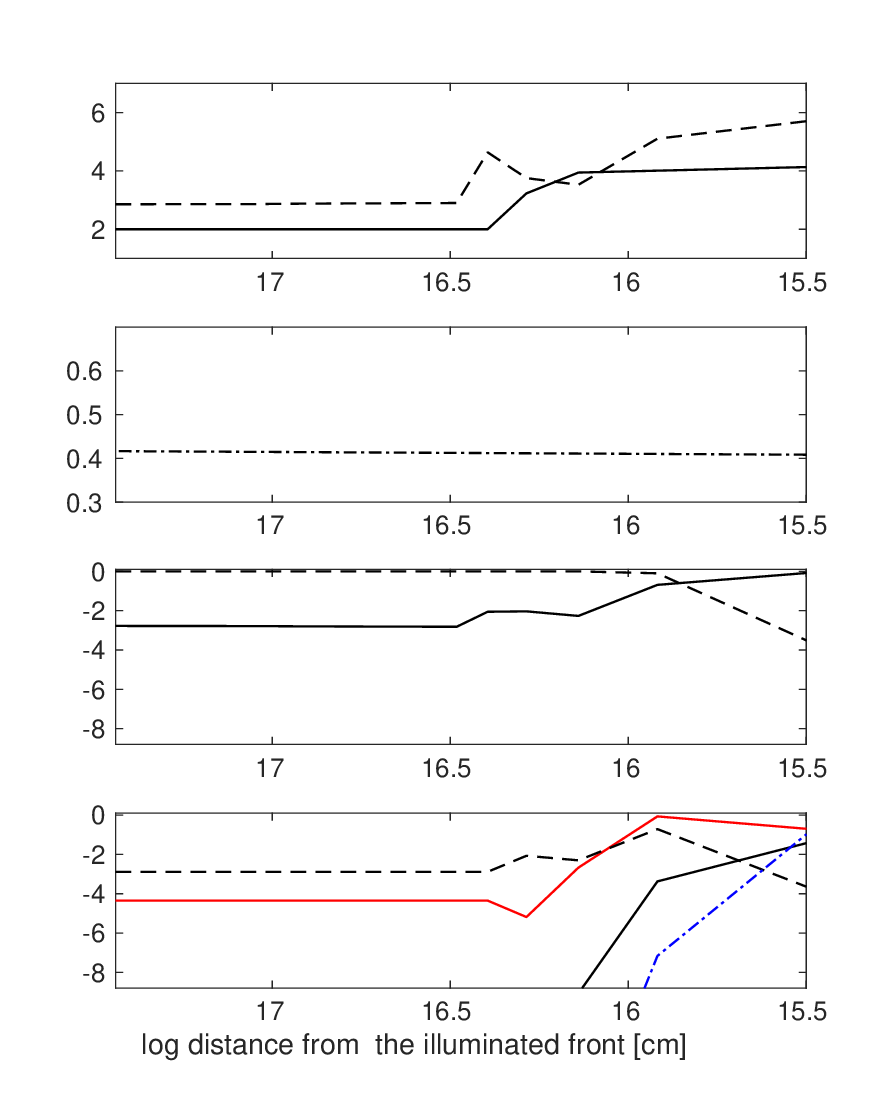}
\caption{Calculated \Te, \Ne  profiles and  fractional abundances of the main  ion profiles throughout the
RXCJ2248-ID galaxy clouds presented  by Topping et al.
The clouds corresponding to a str=1 (ejection) model are divided into two contiguous halves which are
displayed by the left and right panels.  In the left panel
the shock front is on the left and the X-axis scale is logarithmic. In the right panel
the right edge is reached by the flux from the SB. The X-axis scale is logarithmic
in reverse in order to have the same detailed view of the cloud edges.  }

\end{figure*}

\subsection{Modelling the Sanders et al. (2024) spectra}

Sanders et al. (2024) report the detection of the [OIII]4363 line in 16 galaxies at z = 2.1 - 8.7 observed with JWST/NIRSpec as part of the 
Cosmic Evolution Early Release Science (CEERS) survey programme. Through additional measurements, they constructed the first high-redshift \Te-based 
metallicity calibration of
O/H and Ne/H strong-line ratios, valid over the range 12 + log(O/H) = 7.4 - 8.3. They also report the detection of [NII]6585, which enables further 
investigation 
of the N/O-metallicity relation (Contini 2017, 2020).

Sanders et al. discuss how physical properties - including dust reddening, emission-line ratios, electron density and temperature, and oxygen 
abundance - influence 
the results derived for the CEERS sample and other JWST/NIRSpec auroral-line datasets. Their new calibration can be applied to star-forming galaxies 
at z = 2-9, leading to improved metallicity determinations at Cosmic Noon and during the Epoch of Reionisation.
They derived \Te\ and direct oxygen abundances for a combined sample of 46 star-forming galaxies at z = 1.4-8.7. The methods adopted by Sanders et al. 
are described in detail and with considerable precision.

In contrast, we apply a longer but more detailed modelling approach, calculating line-by-line fluxes and comparing them with the observed values 
across a model grid until acceptable agreement is achieved (within 20\% for strong lines and 50\% for weak lines). The modelling results for the 
Sanders et al. data are presented in Fig. 3 and Tables A.8 and A.9. Models marked with a ‘p’ are part of a pluri-cloud configurations.

For the Sanders et al. sample, a rapid increase of N/O over cosmic time on large scales may indicate the presence of dust, possibly implying 
that a fraction of oxygen
is locked into grains. The survival probability of dust grains increases if they are rapidly decelerated and stopped. Atoms and 
molecules - particularly oxygen - may 
then be  trapped into dust grains, leading to an increase in the observed N/O ratio.

\begin{figure}
\centering
\includegraphics[width=9.00cm]{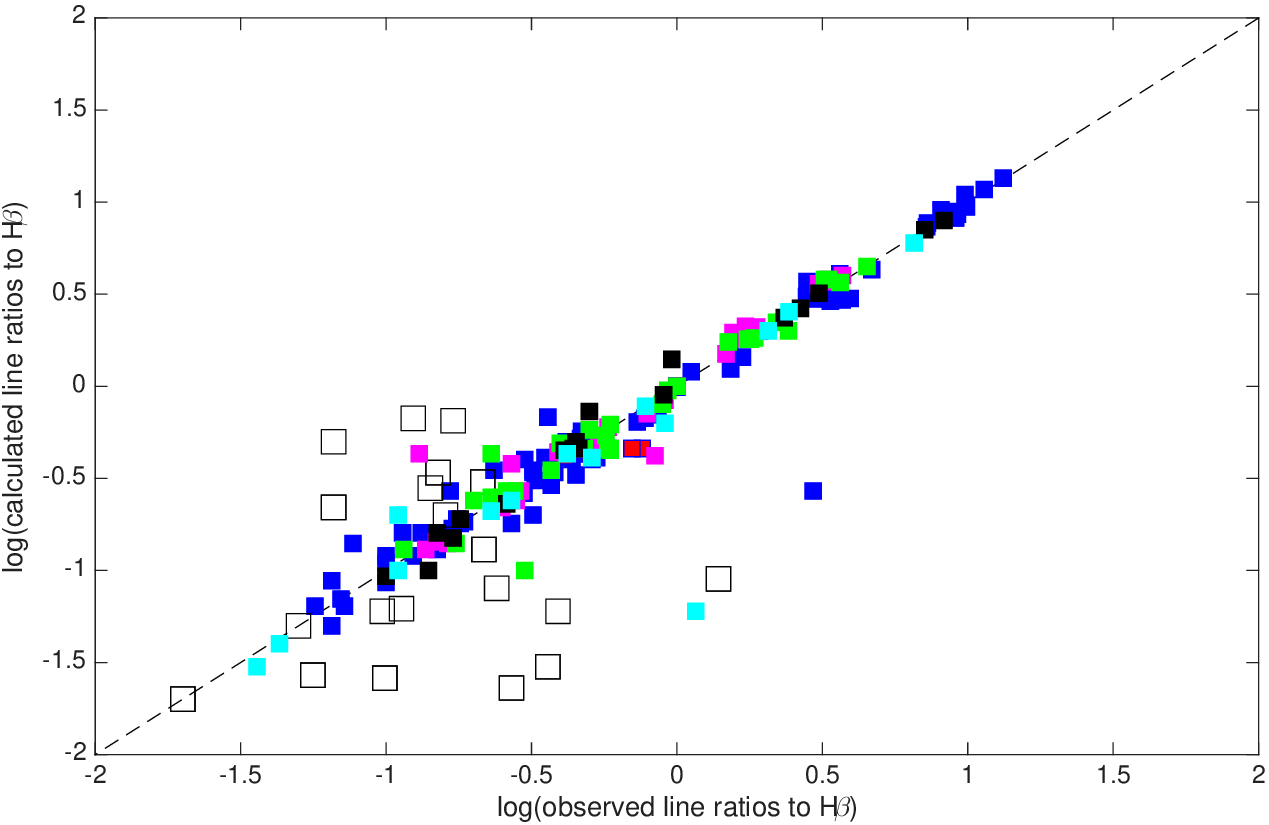}
\caption{Correlations  of calculated versus observed line ratios. Squares: filled blue (Sanders et al); filled red enclosed within blue
(Sanders et al. for \Hg/\Hb $>$ 0.5); empty white large (Sanders et al. for  [SII]6717/\Hb and [SII]6730/\Hb); filled black (Curti et al.);
magenta (Trump et al. on visits 7  except ID4590); filled green  (Trump et al. on visits 8); filled  cyan (Topping et al.).
}
\end{figure}

\section{Discussion and concluding remarks}

In the present investigation of high-redshift galaxies, we began by modelling the spectra in order to constrain the physical conditions (Fig. 4) and 
the chemical 
composition of the emitting gas (Fig. 5). We then discuss the origin of the galactic chemical components based on nucleosynthesis arguments, comparing 
the relative element-abundance trends as functions of redshift and oxygen metallicity with previous theoretical interpretations.

The observed O, Ne, and Balmer lines belong to the Trump et al. (2023) spectra at 5.2753 $\leq$ z $\leq$ 8.4957. With regard to Trump et al., we 
had to analyse 
spectra that were not flux-corrected. Nevertheless, we addressed the abnormally high ratios (e.g. \Hg/\Hb = 0.84) as well as the high but still 
acceptable ones 
(e.g. \Hg/\Hb $\leq$0.59) by adopting pluri-cloud models, thus avoiding strong corrections of high-z spectra.

This approach is justified because models accounting for the coupled effects of shocks and photoionisation can heat the gas to very high temperatures. 
Pluri-cloud models were constructed by combining a radiation-dominated model with a shock-dominated one, enabling us to reproduce both the optical line ratios 
typical of the Osterbrock 'norm' and the high-ionisation lines emitted by collisionally heated gas within the same spectrum.

Curti et al. (2023) corrected the spectra of the Trump et al. objects at z = 7.6597, 7.6651, and 8.68, adding nitrogen lines to the analysis. 
They proposed a flux calibration correction leading to \Hg/\Hb and [OIII]4363/\Hb ratios consistent with the Osterbrock norm. 
Topping et al. (2024) extended the line analysis at z = 6.1, including C, He, Mg, and S lines, modelling the RXCJ2248 galaxy. Sanders et al. (2024) 
incorporated [SII] and additional lines for objects at 2.162 $\leq$ z$\leq$  8.68. Their  calibrated sample enriched the dataset with galaxies showing 
high \Hg/\Hb and low [NII]/\Hb and [SII]/\Hb ratios. In particular, Sanders et al. presented 16 spectra across different redshifts that reveal 
smooth and significant trends.

\subsection{Line ratios}

\begin{figure*}
\centering
\includegraphics[width=9.0cm]{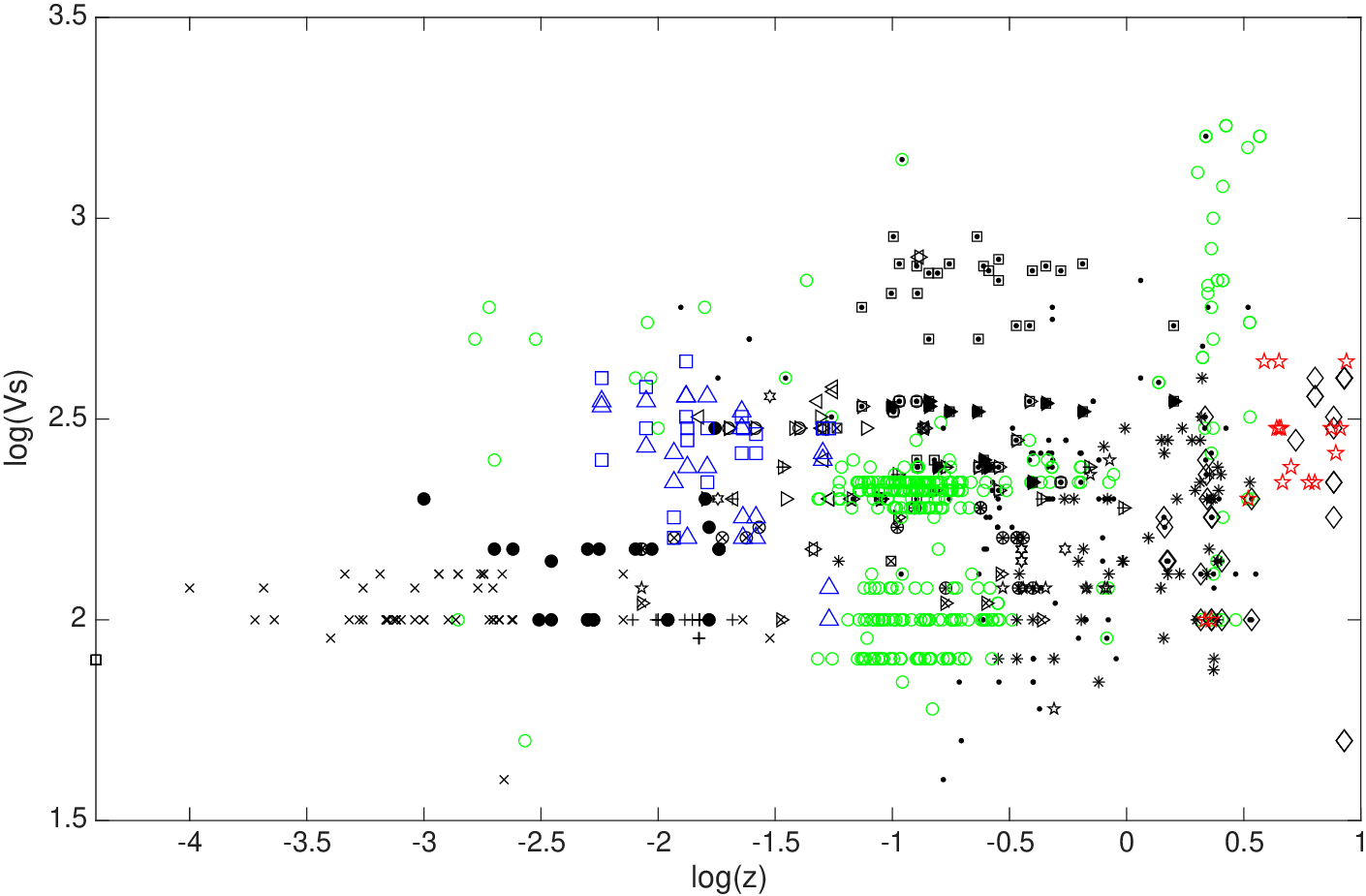}
\includegraphics[width=9.0cm]{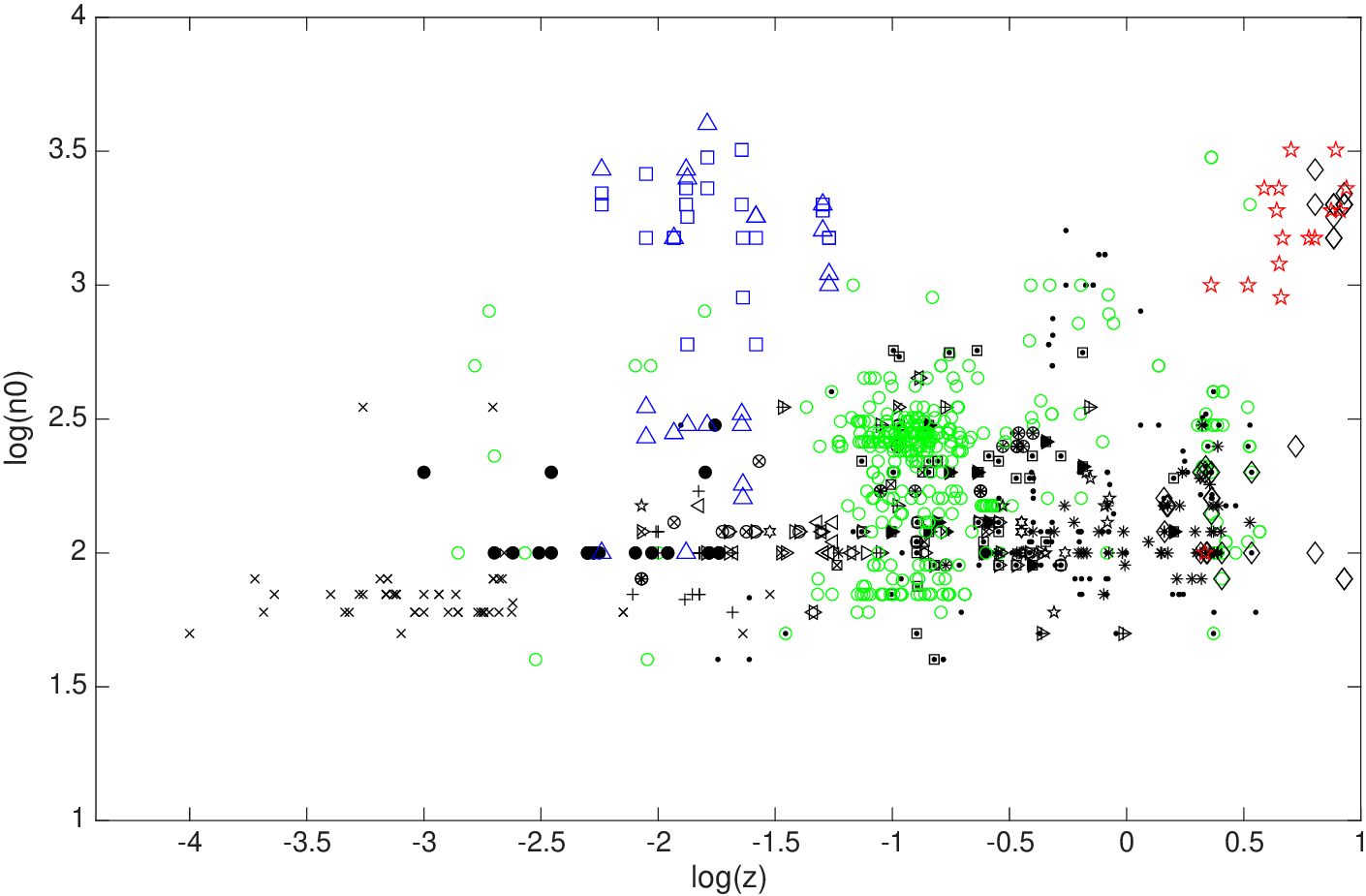}
\caption{Trends of log(\Vs in \kms) and of log(\n0 in \cm3) with the redshift. Symbols are explained in Table 10}

\includegraphics[width=9.0cm]{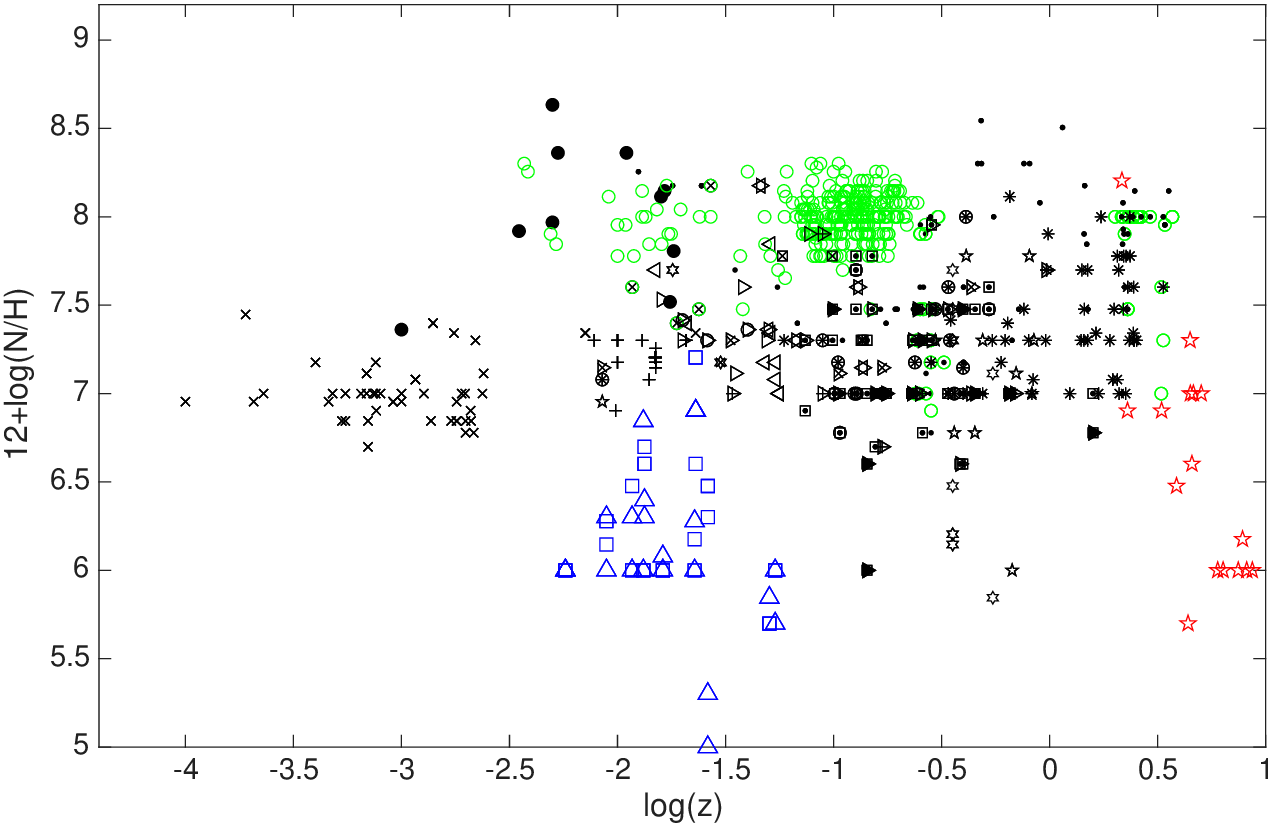}
\includegraphics[width=9.0cm]{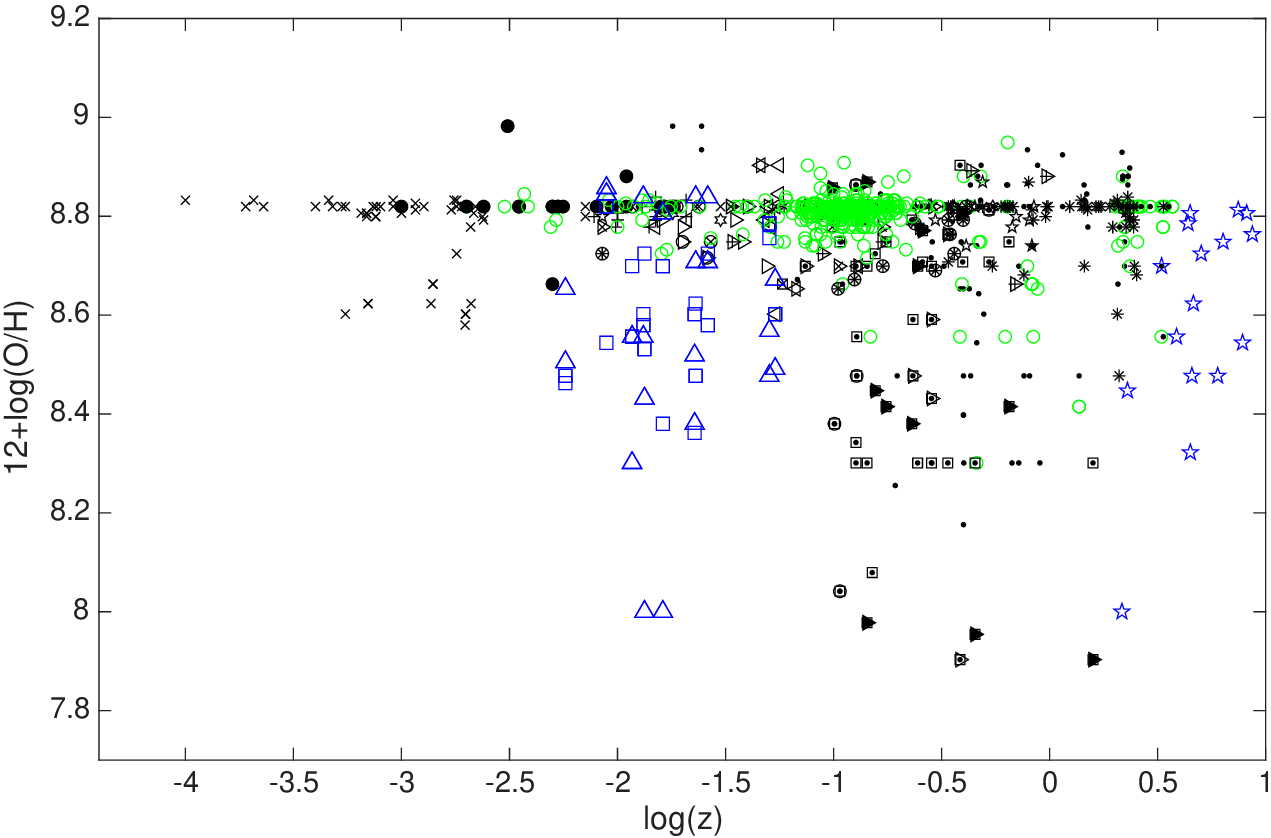}
\includegraphics[width=9.0cm]{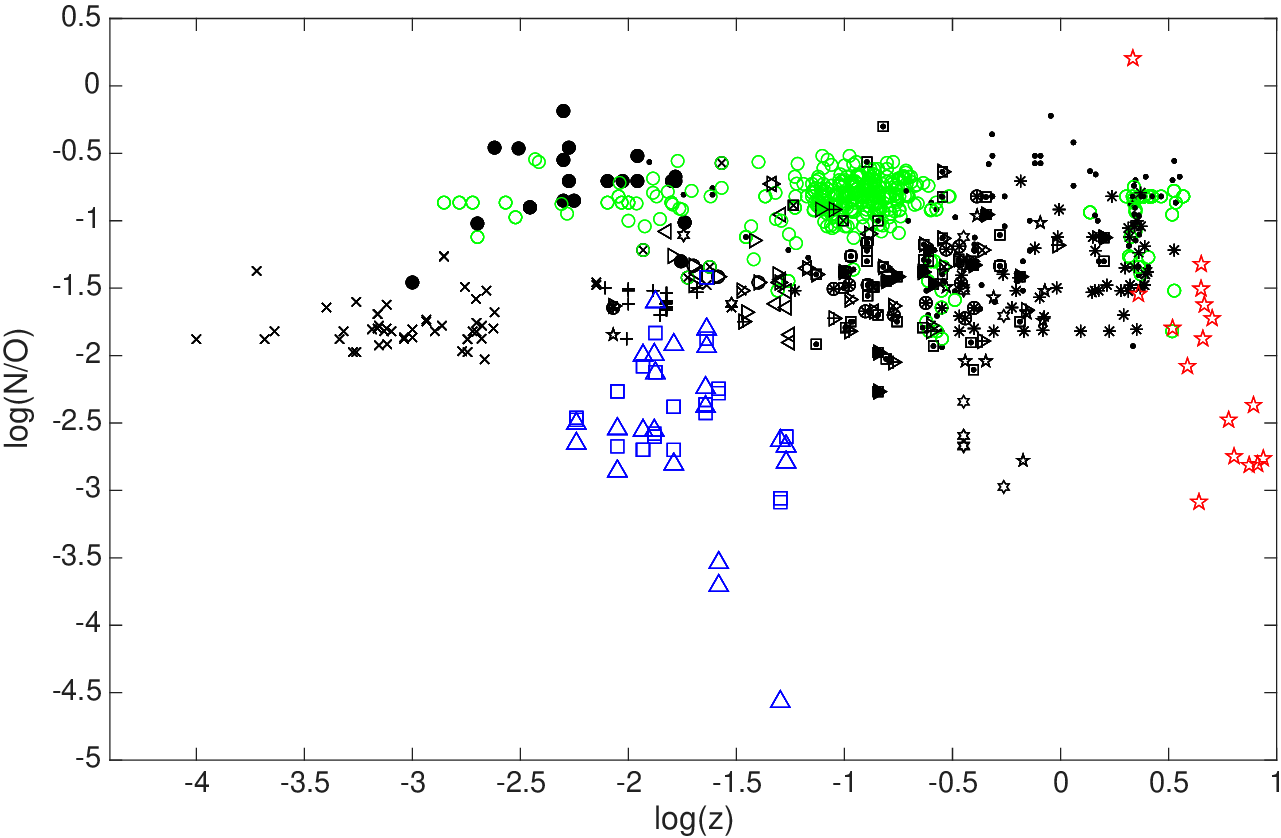}
\caption{Trends of log(N/H), of log(O/H) and log(N/O) with  redshift. Symbols as in Fig. 4}
\end{figure*}

\begin{figure*}
\includegraphics[width=9.0cm]{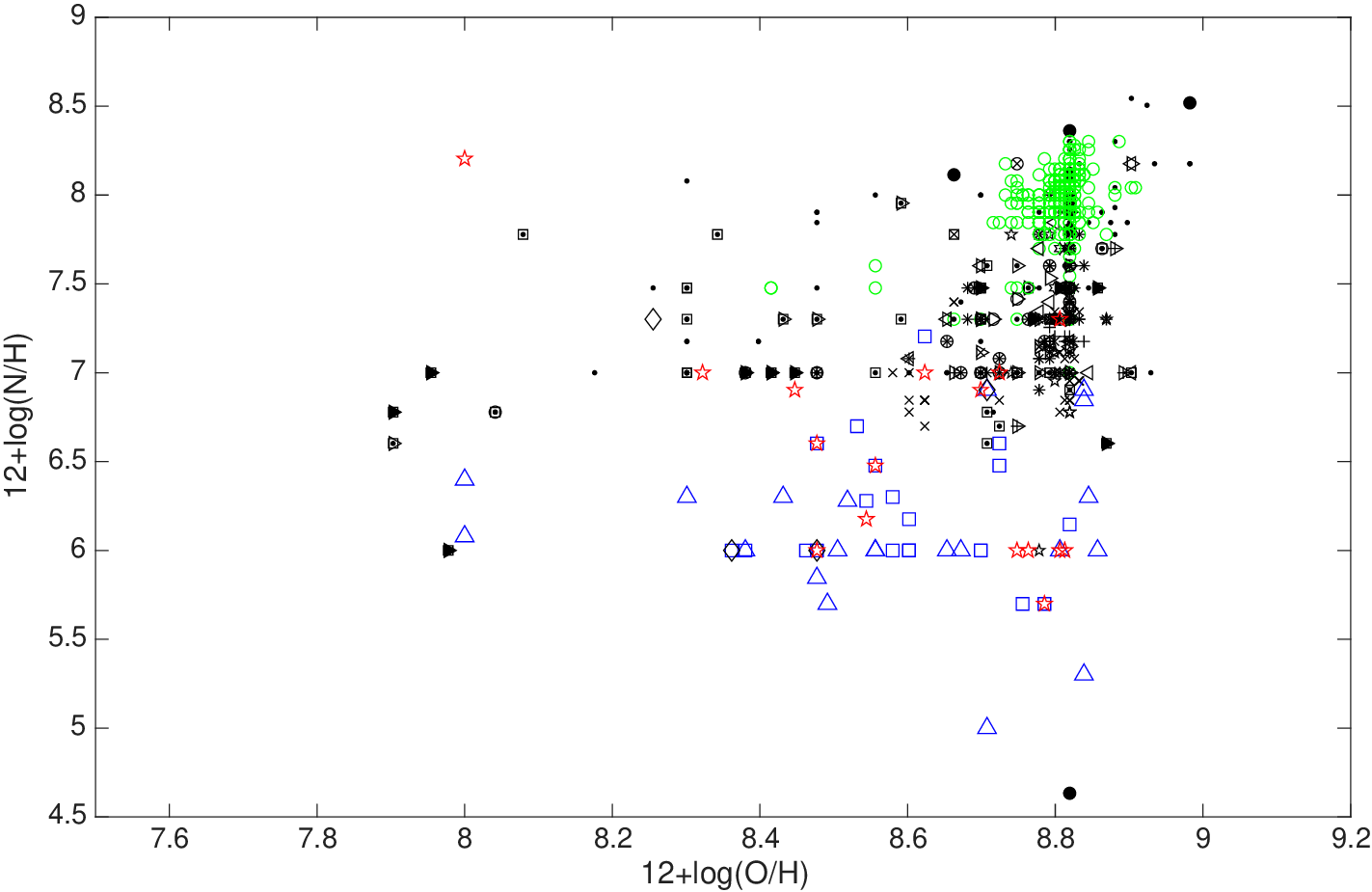}
\includegraphics[width=9.0cm]{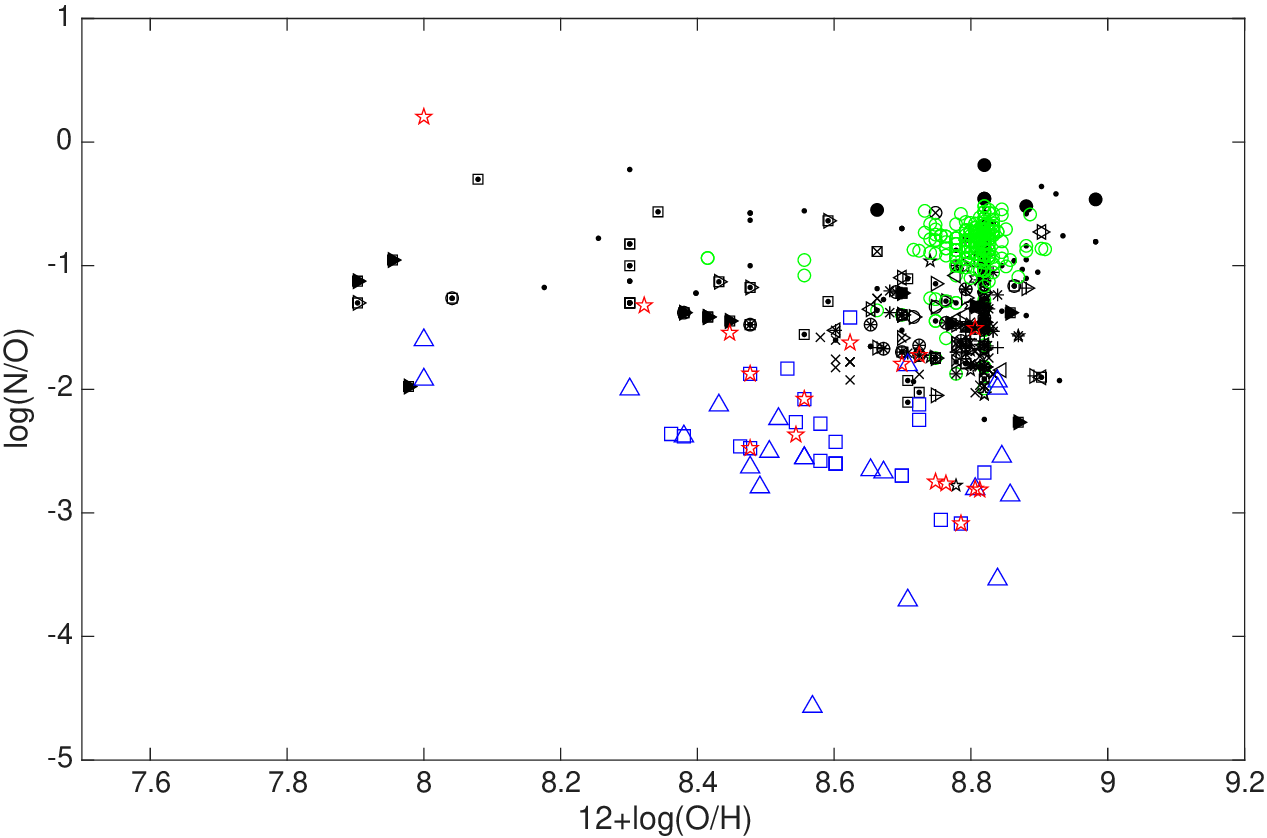}
\caption{Trends of log(N/H) and of log(N/O) with   oxygen metallicity (right). Symbols as in Fig.4}
\end{figure*}

\begin{table*}
\centering
\caption{Symbols in Figs. 4-6}
\begin{tabular}{llcl} \hline  \hline
\ symbol   &object                       & Ref. \\ \hline
\ black encircled dot&  SLSNR hosts           & (1) \\
\  black open triangles at z$\geq$0.1 & SLSNII hosts & (2)\\
\  black filled triangles & SLSNI hosts         & (3)\\
\  black square+dot & only shock models for SLSN hosts & (4)\\
\  black square+cross& Type Ic host central     & (5)\\
\  black circle+cross & Type Ic host at SN positions & (6) \\
\  black triangles  at z$\leq$ 0.1 & SN Ib host & (7)    \\
\  black encircled triangles  & SN IIb hosts    & (8) \\
\  black opposite triangles    & SN Ic hosts    & (9) \\
\  black double triangles      & SN IcBL hosts  & (10) \\
\  black hexagrams        & SN Ibc        & (11)  \\
\  black asterisks &GRB hosts                   & (12) \\
\  black triangle+cross    & LGRB hosts         & (13)  \\
\  black pentagrams  & LGRB different hosts     &  (14)  \\
\  black triangle +plus& LGRB hosts with WR stars& (15) \\
\  black encircled asterisks &LGRB at low z     & (16)\\
\  black hexagrams  & SGRB hosts               & (17)\\
\  black dots & starburst galaxies (SB)        & (19,25)\\
\  green open circles & AGN                     & (20,21,25,26,27)  \\
\  black filled circles & LINER                 & (22) \\
\  black plus &  low-luminosity nearby galaxies & (23)  \\
\  black cross& HII regions in local galaxies   & (24)  \\
\  blue open triangles& AGN (EMPG)              & (28)  \\
\  blue open squares&  SB (EMPG)                & (28)  \\
\  black open diamant&  galaxies at high z      &( 29, 30, 31) \\
\  red pentagram    & N/H and N/O in galaxies at high z & (32)\\
\  blue pentagram   & O/H in galaxies at high z & (32)\\ \hline
\end{tabular}

 (1), (2), (3), (4) (Leloudas et al. 2015);
(5), (6) (Modjaz et al. 2008);
(7), (8), (9), (10), (11) (Sanders et al. 2012);
(12) (Kr\"{u}hler et al. 2015);
(13) (Savaglio et al. 2009);
(14) (Contini 2016, table 8);
(15) (Han et al. 2010);
(16) (Niino et al. 2016);
(17) (de Ugarte Postigo et al. 2014);
(19) (20) (Contini 2014);
(21) (Koski 1978, Cohen 1983, Kraemer et al. 1994);
(22) (Contini 1997);
(23) (Marino et al. 2013);
(24) (Berg et al. 2012);
(25) (Contini 2016);
(26) (Viegas \& Contini, M. 1994);
(27) (Dors et al. 2021);
(28) (Nakajima et al. 2022);
(29) (Trump et al. 2023);
(30) (Curti et al. 2023);
(31) (Topping et al. 2024);
(32) (Sanders et al. 2024).
This table  was updated from Contini (2017, Table 9).

\end{table*}

All spectra were modelled including the coupled effects of photoionisation and shocks.

The agreement between calculated and observed line ratios is shown in Fig. 3, except for [SII]/\Hb ratios, where the relative uncertainty is 
larger due both to ISM contributions and to the lower elemental abundances.

Nearly all \Hg/\Hb ratios reported by Trump et al. range between 0.54 and 0.59, about 15\%
 higher than typically observed at lower redshift ($\leq$0.4 to $\leq$0.5). The abnormally high \Hg/\Hb value in ID4590 (visit 7) led Curti et al. 
 to suspect a shutter malfunction. They, therefore, analysed visit 8 and presented revised modelling for ID4590, ID6355, and ID10612.
Curti et al. derived, using the direct method, 12+log(O/H) = 7.39-7.50 (ID4590), 7.76-7.86 (ID6355), and 7.91-7.88 (ID10612). Our modelling yields 
higher values:
8.176-7.95 (ID4590, visits 7-8),
8.56-8.53 (ID6355),
8.4-8.4 (ID10612).
The discrepancy can be explained by shock heating and ionisation effects in the emitting clouds.

Flux-calibrated spectra with 0.41 $ < $ \Hg/\Hb $ < $ 0.76 were published by Sanders et al. (2024), while de Ugarte Postigo et al. (2014) reported uncorrected spectra with 
\Hg/\Hb = 0.77. The coexistence of gas under different physical conditions within galaxies motivates modelling without applying strong corrections, 
since both low- and high-temperature line ratios may be emitted simultaneously.
In local metal-poor galaxies (e.g. Nakajima et al.), \Hg/\Hb $\sim$ 0.44-0.49 at most. Ratios $>$0.5 in some Sanders et al. high-z spectra support 
pluri-cloud modelling without invoking strong extinction corrections.

Pluri-cloud models were therefore adopted, summing weighted optical-UV emission from clouds and emission from very hot fragments. 
We assume cloud fragmentation at the 
shock front and downstream. The fragment geometrical thickness occupy the region between the shock front and the recombination-driven temperature drop.
They correspond to optical line ratios approaching zero.

Extreme fragments reproducing \Hg/\Hb  $\geq$0.7 (e.g. Sanders IDs 792 and 1559) require \Vs $\sim$ 8300 \kms, \n0 $\sim$ 10$^7$ \cm3, and  
$D$ = 2.5$\times$10$^{-7}$ pc (Table A.9). 
More moderate cases (e.g. Topping et al.) with \Hg/\Hb $\geq$ 0.51 are reproduced by \Vs $\sim$ 200 \kms, \n0 $\sim$ 6000 \cm3, and 
$D$ = 2.87$\times$10$^{-5}$ pc.

Modelling suggests that the high \Hg/\Hb values observed in cosmic-dawn galaxies (Trump et al.) - compared with local extremely metal-poor galaxies 
(EMPGs), which 
rarely exceed \Hg/\Hb = 0.50 - can be explained by fragment destruction over the long cosmic time interval between high and low redshift.

\subsection{Shock parameters}

From previous investigations (Contini, 2026b and references therein), we compiled model results for shock velocity (\Vs), preshock density (\n0), 
and N/O ratios derived from N/H and O/H for AGN, starbursts (SB), SN hosts, long and short GRB hosts, and other galaxy types (Table 10). 
We added the present results for cosmic-dawn galaxies and Nakajima et al. EMPGs for comparison.

A characteristic feature of both high-z galaxies and EMPGs is the relatively high preshock density (Fig. 4, right panel). 
Local EMPGs show high \n0 and moderate O/H depletion similar to cosmic-dawn galaxies. Notably, EMPGs at z = 0.005-0.05 exhibit densities comparable 
to those at z $>$ 3. 
This suggests that remnants of pristine galaxies may survive at low redshift, e.g. HSC J1411-0032 and SDSS J1044+0353 (Contini 2026b).

High \n0 accelerates the downstream cooling  
rates until temperatures decrease from $>$10$^5$ K to $\sim$10$^4$ K and below, transitioning from highly ionised to recombined gas.
Star formation begins when densities reach $\geq$3$\times$10$^4$ \cm3. Even clouds with \n0 $\leq$ 10$^4$ \cm3 can reach high densities downstream 
due to compression. 
The clouds are matter-bounded with geometrical thickness $D$$\sim$ 10$^{-4}$-10$^{-5}$ pc and emit high [OIII]5007+/[OII]3727+ ratios. 
Starburst spectra dominate, though AGN contributions cannot be excluded.
Preshock density increases with redshift toward earlier cosmic times. Since star formation rate scales with ISM density (Mouhcine \& Contini 2002),
higher densities at early epochs are expected.

Fig. 4 shows \Vs~ and \n0 as functions of redshift. A weak but clear decreasing trend of both parameters toward lower z is found, consistent with 
Topping et al. (2024). With regard to the preshock density, \n0 increases with z toward early times, approaching the Big Bang.
 Moreover, for \n0, EMPG preshock densities, similar to those of high-z galaxies, emerge from the decreasing trend of the upper edge, suggesting a pristine 
origin of  the galaxies  within the  0.0057-0.0537 z range.
For \Vs, whether the decreasing trend is physical or due to limited statistics,  remains uncertain.
Shock velocities, produced by e.g. galaxy mergers and starburst activity, are expected to be stochastic rather than smoothly evolving.

\subsection{N/O trends}

N/H, O/H, and N/O were derived from shock+photoionisation models for many galaxy types. Figs. 5 and 6 display these ratios versus redshift and metallicity, 
respectively.

If ${\alpha}$-capture and CNO-cycle processes dominated, N/O trends would scale with He/H and C/H. However, high He/H ratios are not generally observed in the 
present high-z sample, and only a few galaxies show solar N/H. Evidence for strong Wolf-Rayet enrichment is limited, except possibly for J082555.

Fig. 5 shows 12+log(N/H) between 6.9-8.2 and 12+log(O/H) between 8.3-8.82. Sanders et al. data suggest a rising N/H with decreasing z,  
 possibly reflecting the 
first dredge-up. At z $\sim$ 0.3, a second dredge-up may be traced by SGRB hosts, reaching the LINER-AGN domain.
The O/H lower data edge  increases toward low z and encompasses the EMPG region. High-z behaviour remains uncertain due to limited data and larger 
uncertainties.
During the thermally pulsing AGB phase (Marigo 2007), third dredge-up processes may enhance N/H. In EMPGs, N/H is depleted by factors $ >$10, while O/H is depleted 
by factors 2-5 relative to solar (12+log(O/H) = 8.39-7.99; solar = 8.69, Grevesse 2019).

In Fig. 6, log(N/H) versus metallicity shows accumulation of galaxy types near solar O/H between first and second dredge-up domains. The flat N/O behaviour at low 
metallicity (Henry et al. 2000) is attributed to low star-formation rates flattening the age-metallicity relation.
EMPG data overlap with Sanders et al. galaxies at similarly low N/H and comparable preshock densities. This strengthens the hypothesis that remnants of pristine 
galaxies may survive at low redshift, providing insight into galaxy conditions at cosmic dawn.

\begin{acknowledgements}
 I am very grateful to the referee for  his constructive criticism which  led to a more detailed presentation of the paper.
\end{acknowledgements}

\appendix
\section{Tables}

\begin{table*}
\centering
\caption{Comparison of  calculated line ratios to \Hb=1 with the observations reported by Trump et al.   on visits 7 and 8 except for ID4590 visit 7.}
\begin{tabular}{ccccccccccc} \hline  \hline
\ ID/z/visit            & [OII]3727+3729/\Hb & [NeIII]3869/\Hb& \Hg/\Hb  & [OIII]4364/\Hb & [OIII]4959/\Hb & [OIII]5007/\Hb  \\ \hline
\ m(fr)*               & 0                  &0               &0.70      &0               &   0            & 0            \\
\ 4590/8.4957/8        &0.174               &0.37            &0.59      &0.23            &  0.5           & 1.5          \\
\ mtr$_{8.49}$(cl8)    &0.14                &0.35            &0.45      &0.25            &  0.58          &1.74          \\
\ mtr$_{8.49}$(fr8)    &0                   &0               &0.57      &0               &  0             &0             \\
\ mtr$_{8.49}$(cl8A)   &0.167               &0.39            &0.44      &0.27            &  0.64          &1.9           \\
\ 6355/7.6651/7        & 0.91               & 0.58           & 0.59     &0.163           &   1.89         & 3.71         \\
\ mtr$_{7.66}$(cl7)    &0.84                & 0.60           &0.46      &0.14            &  2.0           & 4. 2         \\
\ mtr$_{7.66}$(fr7)    & 0                  &0               &0.54      &0               &  0             & 0            \\
\ 6355/7.6651/8        &0.89                &0.59            &0.48      &0.115           &  1.85          & 3.64         \\
\ mtr$_{7.66}$(cl8)    &0.8                 &0.62            &0.46      &0.13            &  1.83          &3.66          \\
\ 10612/7.6597/7       & 0.24               & 0.54           & 0.54     &0.29            &   1.7          & 3.28         \\
\ mtr$_{7.65}$(cl7)    &0.24                & 0.54           &0.46      &0.27            &  1.8           & 3.8          \\
\ mtr$_{7.65}$(fr7)    & 0                  &0               &0.554     &0               &  0             & 0            \\
\ 10612/7.6597/8       &...                 &0.57            &0.59      &0.26            &  1.75          & 3.32         \\
\ mtr$_{7.65}$(cl8)    &0.24                &0.54            &0.46      &0.27            &  1.8           &3.8           \\
\ 5144/6.3792/7        & 0.25               & 0.39           & 0.54     &0.137           &  1.56          & 3.07         \\
\ mtr$_{6.37}$(cl7)    &0.22                & 0.44           &0.47      &0.13            &  1.96          & 3.6          \\
\ mtr$_{6.37}$(fr7)     & 0                  &0               &0.56      &0               &  0             & 0            \\
\ 5144/6.3792/8        &0.2                 &0.397           &0.58      &0.277           &  1.79          & 3.22         \\
\ mtr$_{6.37}$(cl8)    &0.24                &0.49            &0.46      &0.27            &  1.8           &3.8           \\
\ mtr$_{6.37}$(fr8)    &0                   &0               &0.57      &0               &  0             &0             \\
\ 8140/5.2753/7        & 1.475              & 0.40           & 0.13     & ...            &  1.88          & 3.44        \\
\ mtr$_{5.27}$(cl7)    &1.5                 & 0.48           &0.43      &0.09            &  2.1           & 3.7         \\
\ 8140/5.2753/8        &2.42                &0.93            &0.23      &0.3             &  2.2           & 4.5         \\
\ mtr$_{5.27}$(cl8)    &2.0                 &0.95            &0.43      &0.1             &  2.23          &4.47         \\ \hline
\end{tabular}

*  spectrum calculated to obtain \Hg/\Hb=0.7

\end{table*}

 \begin{table*}
\centering
 \caption{Model input parameters for  Trump et al. spectra (see Table 1).}
\begin{tabular}{llcccccccccccc} \hline  \hline
\ models           & \Vs & \n0  & U    &\Ts      &$F$   &$D$    & O/H      &Ne/H    &str   & W$_{fr}$/W$_{cl}$  &\Hb calc\\
\ units             & 1   & 2    & -    &3        & 4    &  5    &    6     & 6      & -    &        -                  & 7 \\ \hline
\ m(fr)*            &8300 &1e7   & -    & -       & -    & 2.5e-7 & 1.5      &0.15   & 1    &-                           &0.012\\
\ mtr$_{8.49}$(cl8)  &400 &2000  &0.006 &   5     &   -  &1.8e-4  &0.8       &0.12   &0     & -                          &0.014\\
\ mtr$_{8.49}$(fr8)  &400 &2000  &  -   &    -    &   -  &1.e-5   &0.9      & 0.12   & 0    & 11                   &7.0e-7\\
\ mtr$_{8.49}$(cl8A) &400 &2000  &  -    &   -    & 4.8  &1.4e-4  &0.9      & 0.1    & 0    & -                          &0.014\\
\ mtr$_{7.66}$(cl7)  &220 &1500  &   -     & -    & 10    &3.25e-4&3.6      & 0.3    & 0    & -                          &0.049 \\
\ mtr$_{7.66}$(fr7)  &220 &1500  &  -      &  -   & 10    &8e-5   &3.6      & 0.3    & -    & 13                   &9.2e-6\\
\ mtr$_{7.66}$(cl8)  &220 &1500  &  -      &  -   & 10    &3.2e-4 &3.4      & 0.3    &0     & -                          &0.05  \\
\ mtr$_{7.65}$(cl7)  &300 &2000  &  0.055  & 6.8  & 0     &1.5e-4 &2.5      & 0.22   &0     & -                          &0.036 \\
\ mtr$_{7.65}$(fr7)  &300 &2000  & 0.055   & 6.8  & 0     &2.e-6  &2.5      & 0.22   & 0    &  5.7                    &2.3e-7 \\
\ mtr$_{7.65}$(cl8)  &300 &2000  & 0.055   & 6.8  & 0     &1.5e-4 &2.5      & 0.22   & 0    & -                          &0.036  \\
\ mtr$_{6.37}$(cl7)  &360 &2700  &   -     &   -  & 58    &2.3e-4 &3.6     & 0.2     &  0   & -                          &0.43   \\
\ mtr$_{6.37}$(fr7)  &360 &2700  &   -     &   -  & 58    &7.4e-5 &3.6     & 0.2     & 0    &   3.5                   &1.37e-5 \\
\ mtr$_{6.37}$(cl8)  &300 &2000  &  0.055  & 6.8  & 0     &1.5e-4 &2.5      & 0.2    & 0    &  -                         &0.036 \\
\ mtr$_{6.37}$(fr8)  &400 &2000  & -       &   -  & -     &1.0e-5 &0.9      & 0.12   & 0    & 10                   &7.0e-7\\
\ mtr$_{5.27}$(cl7)  &280 &250   & -       & -    &90     &0.03  &1.8       &0.2     & 1    &4                           &0.5   \\
\ mtr$_{5.27}$(cl8)  &280 &250   & -       & -    & 90    &0.03  &2.25      &0.4     & 1    & 5                          &0.49   \\ \hline
\end{tabular}

 \   *: see Table 1;  1: \kms; 2: \cm3; 3: 10$^4$K; 4: 10$^{10}$ photon cm$^{-2}$ s$^{-1}$ eV$^{-1}$ at the Lyman limit; 5: parsec; 6: 10$^{-4}$; 7: \erg.
\end{table*}

\begin{table*}
\tiny{	
\caption{Comparison of calculated line ratios to \Hb=1 with the observations reported by  Curti et al. (2023) for  ID4590, ID6355 and ID10612  
	 flux-calibrated spectra at z=8.4957, 7.6651 and 7.6597, respectively}
\begin{tabular}{lccccccccccccc} \hline  \hline
\  ID/z               &CIV/\Hb  &HeII/\Hb &OIII]/\Hb&CIII]/\Hb&[OII]/\Hb &[NeIII]/\Hb  &\Hg/\Hb &  [OIII]/\Hb &[OIII]/\Hb &[OIII]/\Hb &[NII]/\Hb \\
\                     &   1540  &    1640 &     1663&     1907&     3727+&       3869  & 4340   &4363         &    4959   &      5007 &      6584 \\ \hline
\ 4590$^1$            &  -      &    -    &   -     &   -     & 0.15     &0.17         &0.41    & 0.14        & 0.96      &  3.08     &    -       \\
\ J082555$^2$         &0.97     & 0.36    & 1.3     & 6.1     & -        &0.276        &0.476   & 0.116       & 1.232     &  3.62     & 0.014      \\
\ mcr(B1m)$^3$       &1.46     & 0.28    & 0.84    & 8.8     &0.156     &0.14         &0.46    & 0.14        & 1.37      &  4.1      & 0.017      \\
\ mcr$_{8.49}$(cl)$^4$ & -       &   -     &   -     &   -     &0.16      &0.30         &0.45    & 0.075       & 1.4       &  3.2      & 0.014      \\
\ 6355$^1$            &    -    &   -     &  -      &  -      & 0.90     &0.45         &0.46    & 0.10        & 2.66      &  8.29     &  -        \\
\ mcr$_{7.66}$(cl)$^5$  &         &         &         &         & 0.90     &0.5          &0.46    & 0.093       & 2.65      &  7.95     & 0.18      \\
\ 10612$^1$           &  -      &   -     &  -      &  -      & 0.26     &0.50         &0.44    & 0.18        & 2.34      & 7.11      &    -      \\
\ mcr$_{7.65}$(cl)$^6$  &         &         &         &         & 0.23     &0.73         &0.46    & 0.19        & 2.36      & 7.08      & 0.013     \\ \hline
\end{tabular}}

	$^1$ observed and    flux-calibrated by  Curti et al. (2022);
$^2$ observed and interpreted by Berg et al. (2016) for J082555;
$^3$  calculated by Contini (2026a, Table 10, model B1m);
$^4$  calculated by Contini (2026b on the basis of Table 7, model  ms9.1);
$^5$  calculated by Contini (2026b on the basis of  Table 5, model mh11.1);
$^6$  calculated by Contini (2026b on the basis of Table 8, model ms12.0).

\end{table*}

\begin{table*}
\centering
\caption{Model input parameters for  Curti et al. (2024) spectra calculated in this paper (see Table 3).}
\begin{tabular}{llcccccccccccccc} \hline  \hline
\ models            & \Vs  & \n0  & U    & F    &\Ts  & D       & He/H &C/H   & N/H &  O/H     &Ne/H    &str   &\Hb obs$^1$  &\Hb calc $^2$\\
\                   &  1   &   2  &  -   & 3    & 4   &5        &   -  &  6   &    6&6         &6       &     &  7           &  8 \\ \hline
\ mcr$_{B1m}$(cl)$^3$ & 50 &  80  &0.14  &  -   & 4.1 &0.0116   & -    &1.3   & 0.2 & 1.8      &0.2     &1     & 230.8       &1.14e-4  \\
\ mcr$_{8.49}$(cl)$^4$ &400 &2200 & 0.29&  -   & 12  &5.3e-3   &0.002 &0.118 &0.01 & 3.0      & 0.18   &1     & -           &5.98     \\
\ mcr$_{7.66}$(cl)$^5$ & 180 &1800 &  -  & 1e12 & -   &$<$9.3e-4&0.002 &1.3   &0.08 & 5.1      & 0.2    & 1    & -           &1.78     \\
\ mcr$_{7.65}$(cl)$^6$ &320  &2000 &0.099& -    & 15  &8.3e-4   &0.001 &1.3   &0.01 & 2.3      & 0.2    & 0    & -           & 0.72    \\ \hline
\end{tabular}

  1: \kms; 2: \cm3; 3: 10$^{10}$ photon cm$^{-2}$ s$^{-1}$ eV$^{-1}$ at the Lyman limit; 4: 10$^4$K; 5: parsec; 6: 10$^{-4}$; 7: for J082555
        in 10$^{-16}$\erg; 8:\erg;

\end{table*}

\begin{table*}
\centering
	\caption{Comparison of  calculated  UV line ratios to \Hb=1 with the observed   corrected spectrum by Topping et al. (2024) for RXCJ2248 at z=6.1}
\begin{tabular}{lcccccccc} \hline  \hline
\ ID      &NIV/\Hb     &CIV/\Hb      &HeII/\Hb   &OIII]/\Hb    &CIII]/\Hb &MgII/\Hb \\
\         & 1483+      &   1540      &  1640     & 1663+       &1907       &2800     \\ \hline
\ RXCJ2248&1.16        &   2.43       &  0.23    & 1.02        & 0.27      &0.11    \\
\ mtop    &0.06        & 2.54         & 0.21     &0.73         & 0.24      &0.2     \\
\ m0top   &0.013       &0.175         & 0.3      &0            &0.003      &  0     \\ \hline
\end{tabular}

\tiny{
\caption{Comparison of  calculated  optical line ratios to \Hb=1 with the observed   corrected spectrum by Topping et al. (2024) for RXCJ2248 at z=6.1}
\begin{tabular}{lccccccccccc} \hline  \hline
\ ID       &[OII]/\Hb  &[NeIII]/\Hb   &\Hg/\Hb    &[OIII]/\Hb  &HeII/\Hb   &[OIII]/\Hb  &[OIII]/\Hb &HI/\Hb &[NII]/\Hb  &[SII]/\Hb   &[SII]/\Hb \\
\          &3727+      & 3869         &4340       &4363        &4686       &4959        & 5007      &5876    & 6584     &6717    & 6730 \\ \hline
\ RXCJ2248 &0.043      & 0.78         & 0.51      & 0.42       &$<$0.036   &2.06        &6.54       & 0.2    &0.06      &$<$0.027&$<$0.027\\
\ mtop     & 0.04      & 0.78         & 0.41      & 0.43       &0.03       &2.0         &6.0        &0.002   &0.066     &0.016   &0.035 \\
\ m0top    & 0         &  0           & 0.509     &2.4e-4      &0.018      &0           &0          &2.4e-4  & 0        & 0      &0  \\ \hline
\end{tabular}}

\tiny{
\caption{Model input parameters for  Topping et al. (2024) spectra calculated at z=6.1}
\begin{tabular}{llcccccccccccccccc} \hline  \hline
\ models                 & \Vs & \n0                    & U  & \Ts   & D    & He/H & C/H   & N/H  &  O/H  &Ne/H &Mg/H  &str  &\Hb obs  &\Hb calc \\
\                        & 1   & 2                      & -  &3      &  4   &  5   &  6     & 6    &    6  &  6  &  6   & -   & 7           &  8     \\ \hline
\ RXC2248 $^9$           &-    &6.4-31.0 $\times$ 10$^4$ &0.1 &-      & -     &  -   &0.16   & 0.11 &0.269  &-    &  -   & -   & 25.4       & -      \\
\ mtop$_{6.1}$(cl) $^{10}$  & 200 &6300                    &6.0  &9     &0.183  & 0.003&0.8    & 0.09 &2.0    & 0.16&0.0016 & 1  & -          &  5     \\
\ mtop$_{6.1}$(fr) $^{10}$  & 200 &6000                    &6.0  &9     &2.87e-5& 0.003&0.8    & 0.09 &2.0    & 0.16&0.0016 & 1  & -          &  4.12e6  \\ \hline
\end{tabular}}

\ units 1: \kms ; 2: \cm3; 3: 10$^4$K ;4: parsec; 5: -; 6: 10$^{-4}$;  7:  from Topping et al. in 10$^{-19}$ \erg; 8:  calculated in this work in \erg;
	\ $^9$ observed and   corrected by Topping et al. (2024); $^{10}$ calculated in this work.

\end{table*}

\begin{table*}
\centering
\tiny{
	\caption{Modelling Sanders et al. (2024)  flux-calibrated line ratios to \Hb=1}
\begin{tabular}{cccccccccccccccc} \hline  \hline
\ ID &   [OII]+ & [NeIII] &\Hg   &[OIII]    & [OIII]+ &\Ha     & [NII] & [SII]  & [SII]  & [OII]+  & \Hb obs    &  \Hb calc       &  z  \\
\    &  /\Hb    &  /\Hb   &/\Hb  &/\Hb     & /\Hb    & /\Hb   & /\Hb  &  /\Hb  &  /\Hb  &  /\Hb    &     -        &    -           & -   \\
\    &    3727+ &  3869   &4340  & 4363    &  5007+  & 6563   &  6585 &  6718  &  6733  &  7320+   &10$^{-18}$\erg &\erg          & -   \\  \hline
\ 1019p  & 0.38  &  0.5    &0.53  & 0.18    &  9.9    & -      &  -    &  -     &   -    &  -        & 10.17      &     -           & 8.68 \\
\ msd1 & 0.34  &  0.5    &0.46  & 0.18    &  9.4    & 3.1    & 0.012 & 0.011  & 0.026  & 1.12      &   -        &   4.06          & -    \\
\ 1149p  &0.776  & 0.36    &0.506 & 0.15    &  9.21   & -      & -     &  -     &   -    &   -       & 1.52       &   -             & 8.175\\
\ msd2 &0.67*  & 0.68    &0.46  & 0.13    &  8.6    & 2.9    & 0.012 & 0.02   & 0.047  & 1.8       & -          &  0.27           & -    \\
\ 1027  &0.186  & 0.466   &0.42  & 0.175   &  8.59   & -      & -     &  -     &   -    &   -       & 17.0       &   -             & 7.819\\
\ msd3 &0.184  & 0.52    &0.46  & 0.19    &  8.89   & 3.07   & 0.012 & 0.003  & 0.012  & 0.7       &   -        &  7.29           & -    \\
\ 698   &0.416  & 0.32    &0.459 & 0.10    &  9.03   & -      &  -    &  -     &   -    &   -       &  12.85     &   -             & 7.470\\
\ msd4 &0.5    & 0.34    &0.46  & 0.12    &  8.23   & 2.9    & 0.013 & 0.019  & 0.045  & 1.78      &   -        &  0.27           & -    \\
\ 792p   &1.53   & $<$0.46 &0.76  &  0.47   &  11.4   & 3      &$<$0.52&$<$0.36 &$<$0.39 & -         &  1.33      &   -             & 6.357\\
\ msd5 &1.24   &0.43     &0.46  & 0.43    &  11.7   & 2.98   &0.07   & 0.03   &0.06    & 1.3       &  -         & 0.0167          & -    \\
\ 397   &0.73   &0.377    &0.44  & 0.114   &  8.8    & 2.94   &0.166  &$<$0.066&$<$0.066&$<$0.86    &  5.59      &   -             & 6.0     \\
\ msd6 &0.64   &0.38     &0.46  & 0.16    &  8.8    & 2.7    &0.27   &   0.22 &   0.5  &   0.7     &      -     &  0.05           & -       \\
\ 1536p  & 0.436 &  0.47   &0.51  & 0.32    &  9.78   & 3.28   &$<$0.24&$<$1.39 &$<$0.16 &$<$0.39    &  2.79      &   -             &5.033  \\
\ msd7  & 0.4   &  0.57   &0.46  &0.20     &  11.0   & 3.15   &0.18   &0.09    &0.2     & $<$0.173  &    -       &  240            & -    \\
\ 1477  & 1.0   & 0.3     &0.427 &0.169    &  8.52   & 3.37   &0.37   &$<$0.099 &$<$0.24& 0.09      &  11.74     &   -             &4.63  \\
\ msd8 & 0.99  & 0.4     &0.46  &0.169    &  8.57   & 2.9    &0.29   &0.026   & 0.08   &$<$0.14    &   -        & 0.05            &-     \\
\ 174p  & 1.12  & 0.5     &0.5   &0.16     &  9.85   & 2.8    &$<$0.15&$<$0.02 &$<$0.097&$<$0.085   & 3.51       &  -              &4.56  \\
\ msd9  &1.2   & 0.54    &0.44  &0.144    &  9.7    & 3.7    &0.14   &0.02    &0.06    &$<$0.09    &   -        & 1.69            & -    \\
\ 1665  &1.51   & 0.477   &0.412 &0.065    &  9.04   & 3.62   &0.352  &0.124   &0.151   &0.027      & 26.79      &   -             &4.482 \\
\ msd10 &1.6    & 0.49    &0.46  &0.088    &  8.4    & 2.98   &0.41   &0.34    &0.67    &$<$4.5     & -          & 0.13            & -    \\
\ 1559p  &0.53   & 0.40    &0.70  &0.298    &  13.24  & 2.79   &$<$0.28&$<$0.05 &$<$0.27 & 1.68      & 2.15       & -               &4.471 \\
\ msd11& 0.41   & 0.46    &0.46  &0.263    &  13.5   & 3.06   &0.26   &0.023   &0.05    & -         &  -         & 3.64            & -    \\
\ 1651  &0.0    & 0.0     &$<$0.49&0.124   &  7.27   & 0.0    & 0.0   &0.0     &0.0     & 0.0       & 3.64       & -               & 4.375\\
\ msd12&0.55   & 0.7     &0.45    &0.12   &  7.7    & 2.9    &0.012  &0.02    &0.05    & 0.0       & -          & 0.28            & -    \\
\ 11728 & 0.26  &  0.41   &0.476 & 0.167   &  7.22   & 2.8    &$<$0.08&$<$0.056&$<$0.113&$<$0.1     & 6.0        & -               & 3.869\\
\ msd13& 0.24  & 0.43    &0.46  & 0.153   &  7.36   & 3.1    &0.04   &0.027   &0.062   &0.086      &  -         & 4.27            & -    \\
\ 11088 & 1.68  & 0.32    &0.44  & 0.057   &  8.09   & 3.94   &0.326  &0.17    &0.215   &0.072      & 26.32      & -               &  3.3 \\
\ msd14&  1.44 & 0.35    &0.46  & 0.064   &  9.08   & 3.0    & 0.308 &0.65    &0.302   &0.064      &  -         & 0.45            &  -   \\
\ 3788  & 1.93  & 0.51    &0.46  & 0.077   &  9.09   & 3.7    & 0.235 &0.142   &0.217   &0.065      & 39.25      & -               &2.295 \\
\ msd15& 2.0   &0.4      &0.46  & 0.14    &  8.2    & 2.94   & 0.35  &0.28    &0.13    &0.05       &  -         & 0.16            & -    \\
\ 3537  & 0.10  &0.37     &0.41  & 0.132   &  4.68   & 3.63   & 0.45  &0.0685  &0.068   &0.105      & 8.32       &  -              &2.162 \\
\ msd16& 0.11  &0.36     &0.42  & 0.16    &  4.3    & 4.08   & 0.33  &0.0087  &0.008   &0.0095     &  -         &0.0126           & -    \\ \hline
\end{tabular} }

\centering
\tiny{
	\caption{Model parameters calculated to explain  Sanders et al. (2024) spectra  (see Table 8)}
\begin{tabular}{llcccccccccccc} \hline  \hline
\ models             & \Vs  & \n0  & U     &\Ts   &$F$   &$D$     & N/H      & O/H    &Ne/H       \\
\                    & 1    & 2    & -     &3     & 4    &  5     &    6     &  6     & 6         \\ \hline
\  msd1             &440  & 2300 & 0.27  &12    &   -  & 3.3    & 0.01     & 5.8    &  0.2         \\
\  msd2             &300  & 1900 & 0.042 &11    &   -  & 1.46   & 0.01     & 6.4    &  0.2         \\
\  msd3             &260  & 3200 & 0.6   &10    &   -  & 3.0    & 0.015    & 3.5    & 0.18          \\
\  msd4             &300  & 1900 & 0.04  &14    &   -  & 1.46   & 0.01     & 6.5    & 0.15         \\
\  msd5             &220  & 1500 &  -    & -    & 9    & 3.94   & 0.01     & 5.6    & 0.2           \\
\  msd6             &220  & 1500 &  -    &  -   & 15   & 300    & 0.01     & 3.0    & 0.15          \\
\  msd7             &240  & 3200 & 0.16  &  13  & -    & 2.7    & 0.1      &5.3     & 0.2           \\
\  msd8             &220  & 1500 & -     &  -   & 12   & 3.26   & 0.1      &4.2     & 0.15          \\
\  msd9             &300  & 900  & 0.3   & 18   & -    & 33     & 0.04     &3.0     & 0.16           \\
\  msd10            &300  & 1200 & 0.024 & 11.2 & -    & 140    & 0.1      &2.1     & 0.16          \\
\  msd11            &440  & 2300 & 0.24  &12    & -    & 2700   & 0.2      &6.4     & 0.16          \\
\  msd12            &300  & 1900 & 0.042 &11    & -    & 1.5    & 0.005    &6.1     & 0.3           \\
\  msd13            &440  & 2300 & 0.27  &12    & -    & 3.1    & 0.03     &3.6     & 0.15          \\
\  msd14            &200  & 1000 &   -   & -    &22    & 17     & 0.08     &5.0     & 0.18          \\
\  msd15            &100  & 1000 &   -   & -    & 9    & 100    & 0.08     &2.8     & 0.18          \\
\  msd16            &100  & 100  &   -   & -    & 10   & 1330   & 1.6      &1.0     & 0.4           \\ \hline

\end{tabular}}

\ Units. 1: \kms; 2: \cm3; 3: 10$^4$K; 4: 10$^{10}$ photon cm$^{-2}$ s$^{-1}$ eV$^{-1}$ at the Lyman limit ; 5: 10$^{-4}$ parsec; 6: 10$^{-4}$
\end{table*}

\end{document}